\documentclass[]{interact}

\usepackage{amssymb}
\usepackage{amsfonts}
\usepackage{amsmath}
\usepackage{graphicx}

\usepackage{epstopdf}
\usepackage{subfigure}

\usepackage[numbers,sort&compress,merge]{natbib}
\bibpunct[, ]{(}{)}{,}{n}{,}{,}

\theoremstyle{plain}

\theoremstyle{definition}

\theoremstyle{remark}

\begin{document}

\title{Sputtering of targets in high-intensity heavy-ion beam experiments}

\author{
\name{R.~N.~Sagaidak\thanks{CONTACT Email: sagaidak@jinr.ru}}
\affil{Flerov Laboratory of Nuclear Reactions, Joint Institute for Nuclear Research,\\ J.-Curie str. 6, 141980 Dubna, Moscow region, Russia}
}

\maketitle

\begin{abstract}
Based on available models and experimental data, the sputtering of target atoms in long-term experiments with intense heavy ion (HI) beams has been considered. The experiments on the synthesis of superheavy nuclei (SHN), which are carried out in laboratories around the world, are examples of such experiments encountering the target atoms sputtering in specific conditions. Rotating wheels with a relatively large target annulus area are used in these experiments to reduce the temperature and radiation loads on the target. Large areas of rotating targets allow one to reduce the sputtering yields of target atoms significantly. The question arises about the reliability of these yields in experiments on the synthesis of SHN with $Z > 118$. These nuclei can be produced with extremely low cross sections, requiring HI beam doses exceeding 10$^{20}$  particles passed through the target to observe several decay events of the thus synthesized SHN. Estimates based on approximations and simulations considered in this work are presented and compared with some data on actinide target sputtering obtained in experiments performed on the synthesis of SHN.
\end{abstract}


\section{Introduction}
\label{intro}

The detailed study of the properties of superheavy nuclei (SHN) produced in the complete fusion reactions induced by the $^{48}$Ca projectile on actinide target nuclei, which lead to $112 \leqslant Z \leqslant 118$ nuclei, as well as experiments on the synthesis of heavier ones with $Z > 118$, imply the use of heavy ion (HI) beams with significantly higher intensities than those used in the discovery experiments \cite{OgaUtRPP15}. The synthesis of SHN with $Z > 118$ is possible with beam particles such as $^{50}$Ti, $^{54}$Cr, and $^{58}$Fe. For SHN formed in the fusion-evaporation reactions with these projectiles, one may expect production cross sections of $\sim$0.05 pb \cite{ZaGr15} or even less. To observe several decay events of these SHN, one should collect a beam dose exceeding 10$^{20}$ particles. This dose can be provided at an intensity of (3--6)$\times$10$^{13}$ particle/s for a reasonable beam time. Such beams are now provided by the high-current cyclotron DC280 commissioned recently at the Flerov Laboratory of Nuclear Reactions (JINR). The intensity of the $^{48}$Ca beam delivered to a physical target is close to the expected one \cite{DmitrDC280}.

Recoil separators are usually used in experiments on the synthesis of SHN and the study of their properties \cite{DullRev}. The targets of thickness within 0.4--0.7 mg/cm$^{2}$, which are prepared on the Ti backing (actinide targets) or on the carbon one (Pb and Bi targets), are exploited in the experiments with these devices. High beam intensities and beam energy losses inside the targets and backings lead to high heating power released inside them.
Using a pulsing beam is one way to reduce a thermal load. In the case of a continuous beam, a pulsing regime can be realized with a rotating target. The effects of target rotation and beam wobbling on target temperature were recently considered in detail \cite{SagaPRAB21}. High heating powers generated inside a target, as well as high beam doses, reduce its durability. One would think that the target and its backing are mainly degraded, along with thermal loads, due to the sputtering of their atoms and radiation damage to their crystal lattices. All the processes are non-independent, but they can be treated separately \cite{YntNick78}.

Sputtering of target atoms under intense HI beams and high-dose irradiation could be estimated using  Monte Carlo simulations within the TRIM code \cite{SRIM}, a variant of cascade collision models (CCMs). Comparing the results of these simulations with the experimental data \cite{Miesk03} for metal targets, which were earlier considered in \cite{Saga20REDS}, showed an entirely satisfactory agreement of the data for backward sputtering from Au and Zr targets in contrast to the data from the Ti target. For the Ti target, TRIM underestimates sputtering yields by a factor of 3 to 5 (depending on HI and its energy). That seems unacceptable for practical applications (estimates of sputtering yields produced by HIs at energies 4–7 MeV/nucleon), and the appropriate approximation was proposed for the respective estimates \cite{Saga20REDS}. As for the sputtering of UO$_{2}$ target atoms, attention may be drawn to the data obtained with the 4.6 and 3.5 MeV/nucleon Sn and U beams \cite{Bouff98}. These data show the exponential dependence of the HI electronic stopping power (ESP) for backward sputtering yields $S_{b}$. Extrapolation of this dependence to the ESP values for HIs gives us the values of $S_{b} \simeq 1-2$ atom/ion for the 5 MeV/nucleon $^{48}$Ca beam, i.e., the values that are three orders of magnitude higher than those given by the TRIM simulations (0.01 atom/ion).

In general, there is a lack of experimental data at high beam energies for transmission sputtering yields $S_{tr}$ relevant to the actinide targets, which forced us to use appropriate estimates for this value. As noted in the review papers [see, for example, \cite{Wein89,AssTouTra07}], despite a lack of data, sputtering in the high energy regime can be marked by two experimental findings: 1) small yields of particles ejected from clean metal crystalline surfaces, which may be close to those calculated in the framework of CCM (Ti is not the case, as was mentioned above), and 2) high sputtering yields from insulating oxide materials, which are much higher than those from metals and strongly related to the ESP for HI. The latter refers to the actinide oxide targets, with which we deal in the experiments on the synthesis of SHN. In the present study, TRIM simulations \cite{SRIM}, as a variant of CCM, have been considered and compared with available experimental data, aiming at their further applications to estimate the sputtering yields from actinide targets. Thermal spike model (TSM) approximations \cite{AssTouTra07,Meins83} and available experimental data have been used to estimate sputtering yields for those cases. The estimates are mainly performed for the $^{48}$Ca beam, which was the most intensive one achieved in the experiments on SHN studies. Sputtering of heavy atoms from actinide oxide and fluoride targets, as those worldwide used in experiments on the synthesis of SHN \cite{OgaUtRPP15,Kindler08,Lommel08}, is mainly considered in the present work. Experimental observations obtained in our works on the synthesis of SHN are also considered and discussed.

\section{Beam energy dependence in sputtering of U targets}
\label{sputtarg}

\begin{figure}[h] 
\vspace{0.0mm}
\begin{center}
\includegraphics[width=0.65\textwidth]{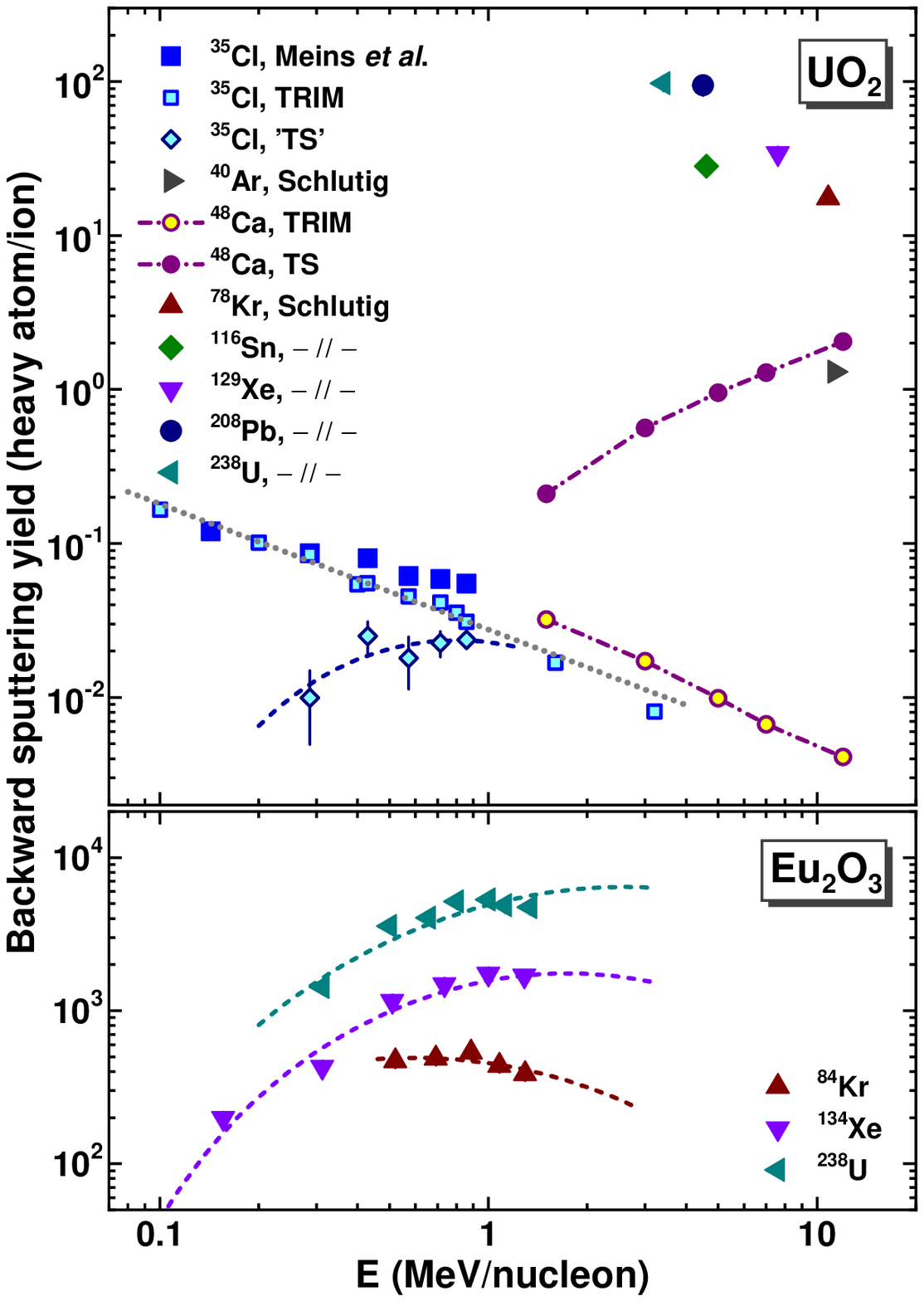}
\vspace{-1.5mm} \caption{Backward sputtering yields of heavy atoms as a function of the HI beam energy for different ions (full symbols) bombarding the UO$_{2}$ \cite{Meins83,Schlut01} and Eu$_{2}$O$_{3}$ \cite{Wein89,Guth86} targets (upper and bottom panels, respectively). The yields of U atoms sputtered by Cl ions \cite{Meins83} are compared to those obtained with TRIM simulations (open squares). The smoothed-out approximation to the TRIM data is shown by a dotted line, whereas the yields of U atoms corresponding to thermal spike (TS) effects are shown by open diamonds. Dashed lines show the results of fitting the data with Eq.~(\ref{TSMenergy}). The sputtering yields of U atoms under the $^{48}$Ca beam, as obtained with TRIM simulations and with the approximation corresponding to the TS effects, are also shown (open and full circles, respectively, connected by dash-dotted lines).}
\label{UErSput}
\end{center}
\end{figure}

One of the first results on UO$_{2}$ sputtering induced by HIs in the MeV energy region was obtained as early as 1983 \cite{Meins83}. The low yields obtained for U atoms sputtered by $^{35}$Cl ions were very close to those obtained with the CCM calculations. In Figure \ref{UErSput}, these data are compared with the results of TRIM simulations. High sputtering yields were later obtained in experiments with heavier ions at higher energies \cite{Bouff98,Schlut01}. These data \cite{Schlut01} are also shown in the figure for comparison.

TRIM simulations for the 1.5--12 MeV/nucleon $^{48}$Ca ions show yields close to the $^{35}$Cl data  \cite{Meins83} extrapolated to higher energies, as shown in the same figure. At the same time, the sputtering yield obtained for the 11.3 MeV/nucleon Ar ion \cite{Schlut01} (another neighboring projectile close to $^{48}$Ca) is about 250 times greater than either extrapolated or simulated values. Much higher sputtering yields were obtained for the Eu$_{2}$O$_{3}$ target \cite{Wein89,Guth86}, as shown in the bottom panel of Figure~\ref{UErSput}.

\begin{figure}[h] 
\vspace{0.0mm}
\begin{center}
\includegraphics[width=0.65\textwidth]{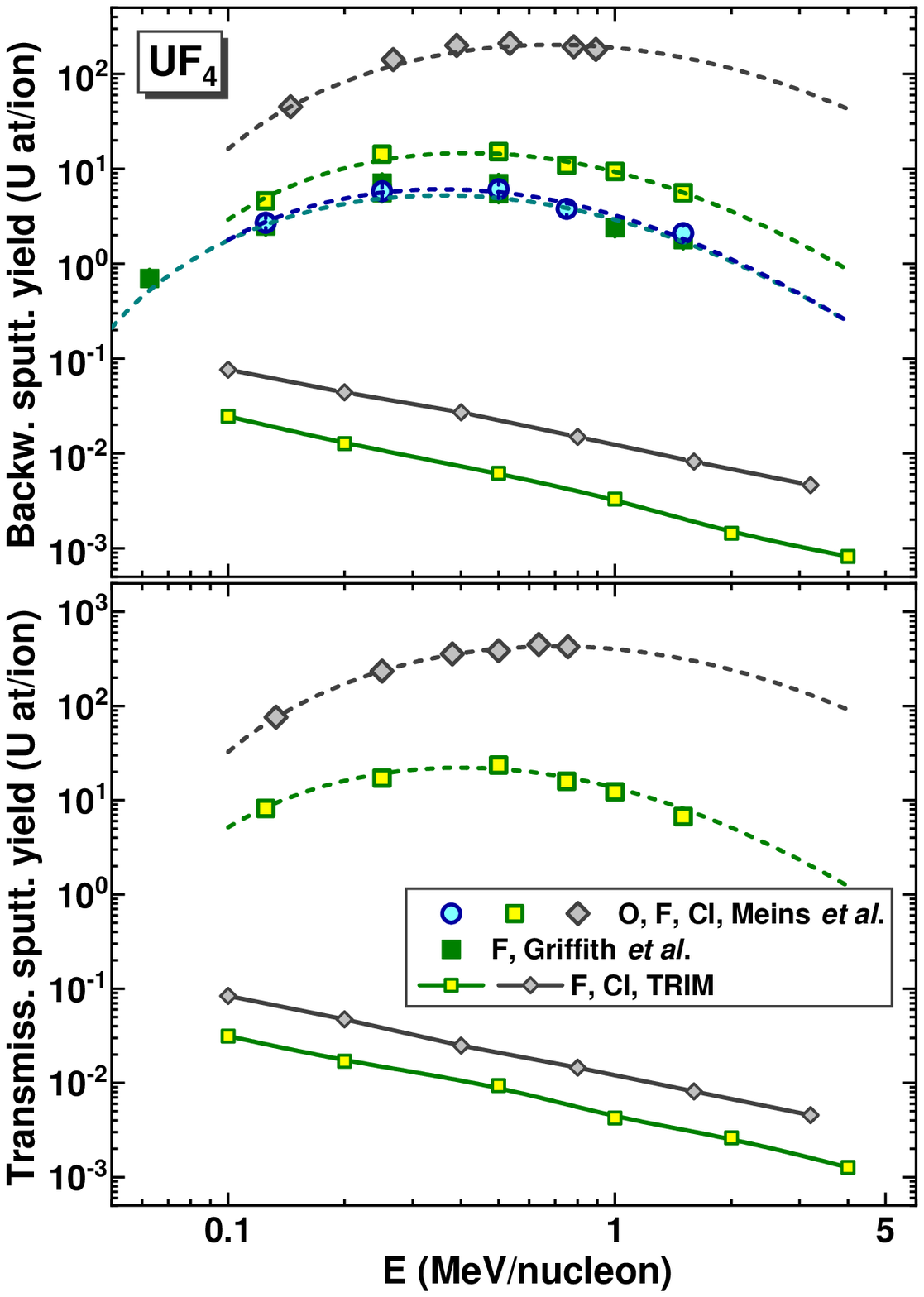}
\vspace{-1.5mm}
\caption{The same as in Figure \ref{UErSput}, but for the yields of U atoms for backward and transmission sputtering (upper and bottom panels, respectively) as the result of the bombardment of the UF$_{4}$ target by different heavy ions (symbols) \cite{Meins83,Griff80}. The data are shown for the equilibrated and non-equilibrated ion charges used in the experiments (open and full symbols, respectively). The results of data fitting with Eq.~(\ref{TSMenergy}) and TRIM simulations for $^{19}$F and $^{35}$Cl beams are also shown (dashed lines and open symbols connected by solid lines, respectively).}
\label{EnerDepU}
\end{center}
\end{figure}

The energy dependence of sputtering yields for Eu atoms, shown in Figure~\ref{UErSput}, and for U ones, sputtered from the UF$_{4}$ target \cite{Meins83,Griff80}, as shown in Figure~\ref{EnerDepU}, are well-fitted with the expression of the fourth power of the HI electronic stopping power. This expression was proposed in the framework of a ``thermalized ion explosion'' model \cite{Seiber80} used in \cite{Meins83} to describe the UF$_{4}$ data. According to \cite{Meins83,Seiber80}, the energy dependence of the observed sputtering yields can be parameterized as
\begin{equation}\label{TSMenergy}
    S \sim [(dE/dX)_{e}]^4 = [A q_{eq}^{2}(E,Z_{\rm HI}) \ln(B E) / E]^4,
\end{equation}
where $(dE/dX)_{e}$ is the ESP of the medium for HI with the atomic number $Z_{\rm HI}$ and energy $E$ (in MeV/nucleon), $q_{eq}(E,Z_{\rm HI})$ in the HI equilibrated charge, which could be calculated according to \cite{Ziegler80}, whereas $A$ and $B$ are the fitting parameters. As will be shown below, sputtering yields correspond to the power function of ESP, but with the index of power differing from 4.

In an attempt to systematize the data for UO$_{2}$ in the framework of the TSM, the thermal spike (TS) component was extracted with the subtraction of the TRIM simulated yields from the total yields obtained in the $^{35}$Cl+UO$_{2}$ experiment \cite{Meins83}. The energy dependence thus obtained for the UO$_{2}$ target sputtering, shown in Figure~\ref{UErSput}, looks like those for UF$_{4}$ and Eu$_{2}$O$_{3}$. Simultaneously, for the last targets, sputtering yields exceed those obtained with TRIM simulations by several orders of magnitude, in  contrast to the UO$_{2}$ target.

The essential findings obtained in these early experimental studies \cite{Meins83,Guth86,Griff80} can be mentioned and used in further applications. Thus, the magnitude of transmission sputtering yields is consistently higher than that of the backward ones, being $\simeq$40\% higher for $^{19}$F and $\simeq$100\% higher for $^{35}$Cl \cite{Meins83} (see Figure~\ref{EnerDepU}). Simultaneously, for the Eu$_{2}$O$_{3}$ target irradiated by U ions, the reverse relationship was observed \cite{Guth86}. For further estimates, it should be considered, bearing in mind that transmission yield is of interest in the application considered here, whereas the most available data refer to backward sputtering. For backward sputtering, the charge state of the HI projectile influences the yield of sputtered atoms. Thus, the yields of U atoms obtained in experiments with the $^{19}$F beam passed through a carbon stripper foil \cite{Meins83} are about two times higher than those obtained with the beam charge states \cite{Griff80} (see Figure \ref{EnerDepU}). The average charges of ions in this case \cite{Shima82} are 10--30\% higher than the equilibrated charges inside the target \cite{Ziegler80}. The lasts determine the ESP and respective sputtering yield. However, the HI beam charge states \cite{Griff80} are 30--55\% lower than the equilibrated charges inside the target. These observations correlate with the dependence of ESP on target thickness, which was observed later \cite{Frey96} and then explained theoretically \cite{Grun05}.

The values of $A$ and $B$ parameters obtained with Eq.~(\ref{TSMenergy}) fitting the data shown in Figures \ref{UErSput} and \ref{EnerDepU} give no way to systematize them because of a lack of data, although any kind of approximation based on these and similar parameter values could be helpful for estimates of the sputtering yields for any HI of specified $Z_{\rm HI}$ and $E$, which interacts with the respective target.

Note that measured yields of U atoms were insensitive to the change in the UF$_{4}$ target temperature within 70--170 $^{\circ}$C, as was observed in the irradiation of the target by 0.5 and 1.0 MeV/nucleon $^{19}$F ions \cite{Griff80}. The temperature of the actinide targets used in experiments on the synthesis of SHN with the intense 5 MeV/nucleon $^{48}$Ca beam may achieve several hundred Celsius degrees, as estimated in \cite{SagaPRAB21}.

In an attempt to systematize the data, the sputtering yields against the ESP for incident ions, which is the crucial value of TSM, were considered, as  it was done in many works [see, for example, \cite{Wein89,AssTouTra07,Bouff98,Schlut01,Guth86,Matsu03,Hedin87}].

\section{Electronic stopping power dependence in sputtering of U targets}
\label{ESPdepSput}

\begin{figure}[h] 
\vspace{0.0mm}
\begin{center}
\includegraphics[width=0.65\textwidth]{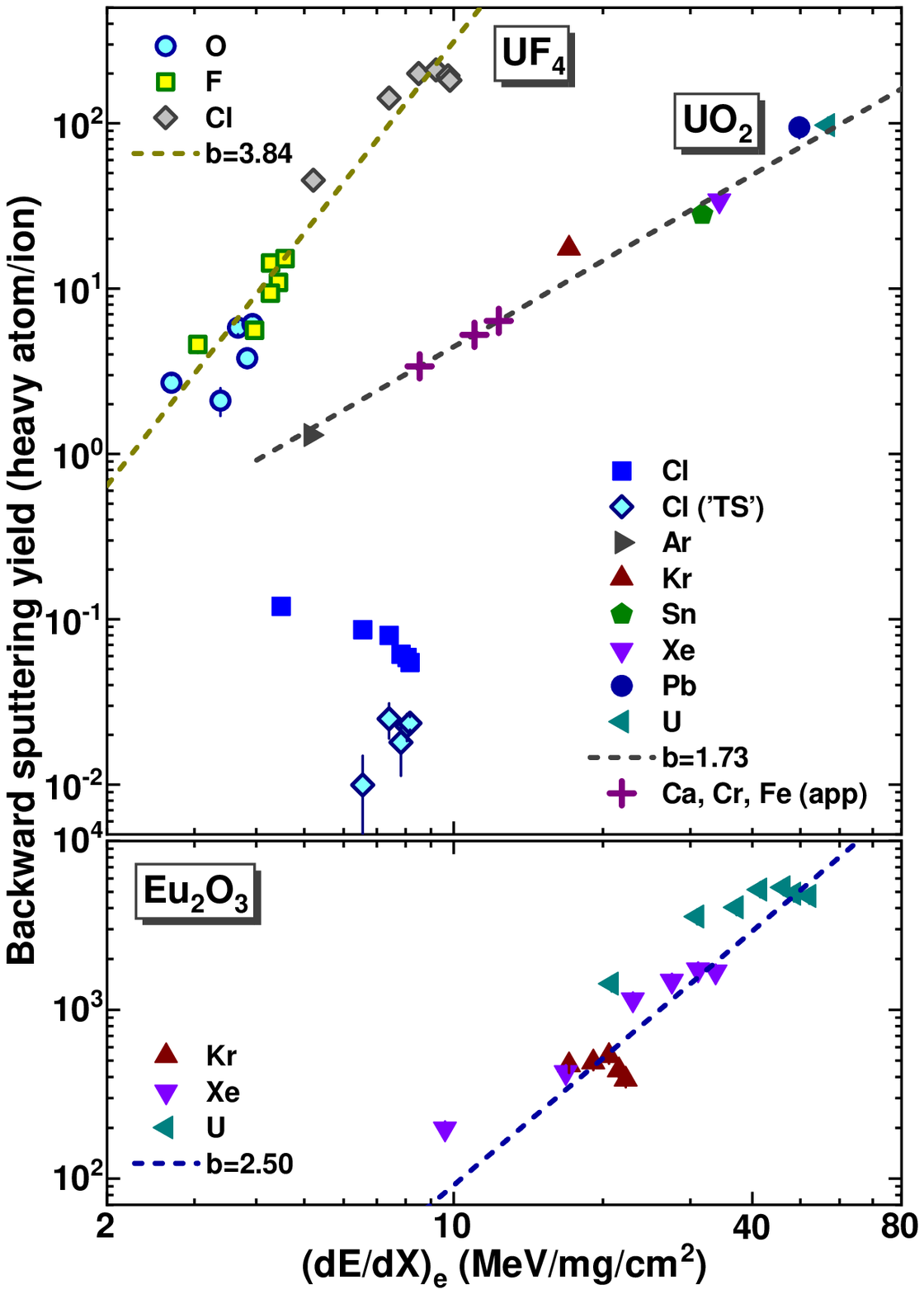}
\vspace{-1.5mm}
\caption{The same data as those shown in Figures \ref{UErSput} and \ref{EnerDepU} for backward sputtering yields \cite{Wein89,Meins83,Schlut01,Guth86,Griff80}, but plotted as a function of electronic stopping powers $(dE/dX)_{e}$ for heavy ions used in experiments (symbols). The results of the $a[(dE/dX)_{e}]^{b}$ function fitting the data for different targets are shown by dashed lines, specified by power index $b$. The sputtering yields for $^{48}$Ca, $^{54}$Cr, and $^{58}$Fe projectiles, as estimated with this approximation, are also shown by crosses.}
\label{SputdEdX}
\end{center}
\end{figure}

In Figure \ref{SputdEdX}, heavy-atom backward sputtering yields obtained in experiments (see Figures \ref{UErSput}  and \ref{EnerDepU}) are plotted for different HIs as a function of the ESP of the respective target media. As one can see, the arrangement of data points does not allow one to make perfect fits with the $a[(dE/dX)_{e}]^{b}$ function ($a$ and $b$ are the fitting parameters) describing the data for specific targets. Exponent values $b$ resulting from such fitting vary from 1.73 to 3.84, as indicated in the figure. Note that sputtering yields of U atoms from UO$_{2}$ under high-energy HIs [from $^{40}$Ar to $^{238}$U \cite{Schlut01}] and $^{35}$Cl \cite{Meins83} (did not use for fitting) contradict each other. It is seen that yield dependencies for UF$_{4}$ \cite{Meins83,Griff80} and Eu$_{2}$O$_{3}$ \cite{Guth86} targets reveal a large spread of data points. The data inconsistency for the UO$_{2}$ target in such a presentation does not allow one to make reasonable predictions for $^{48}$Ca beam energies at 5 MeV/nucleon. Nevertheless, considering the high-energy data only \cite{Schlut01}, the backward sputtering yields obtained with this approximation for $^{48}$Ca, $^{54}$Cr, and $^{58}$Fe projectiles  at respective energies are estimated as 3.4 to 6.4 atoms/ion for UO$_{2}$ targets.

\begin{figure}[h!] 
\vspace{0.0mm}
\begin{center}
\includegraphics[width=0.65\textwidth]{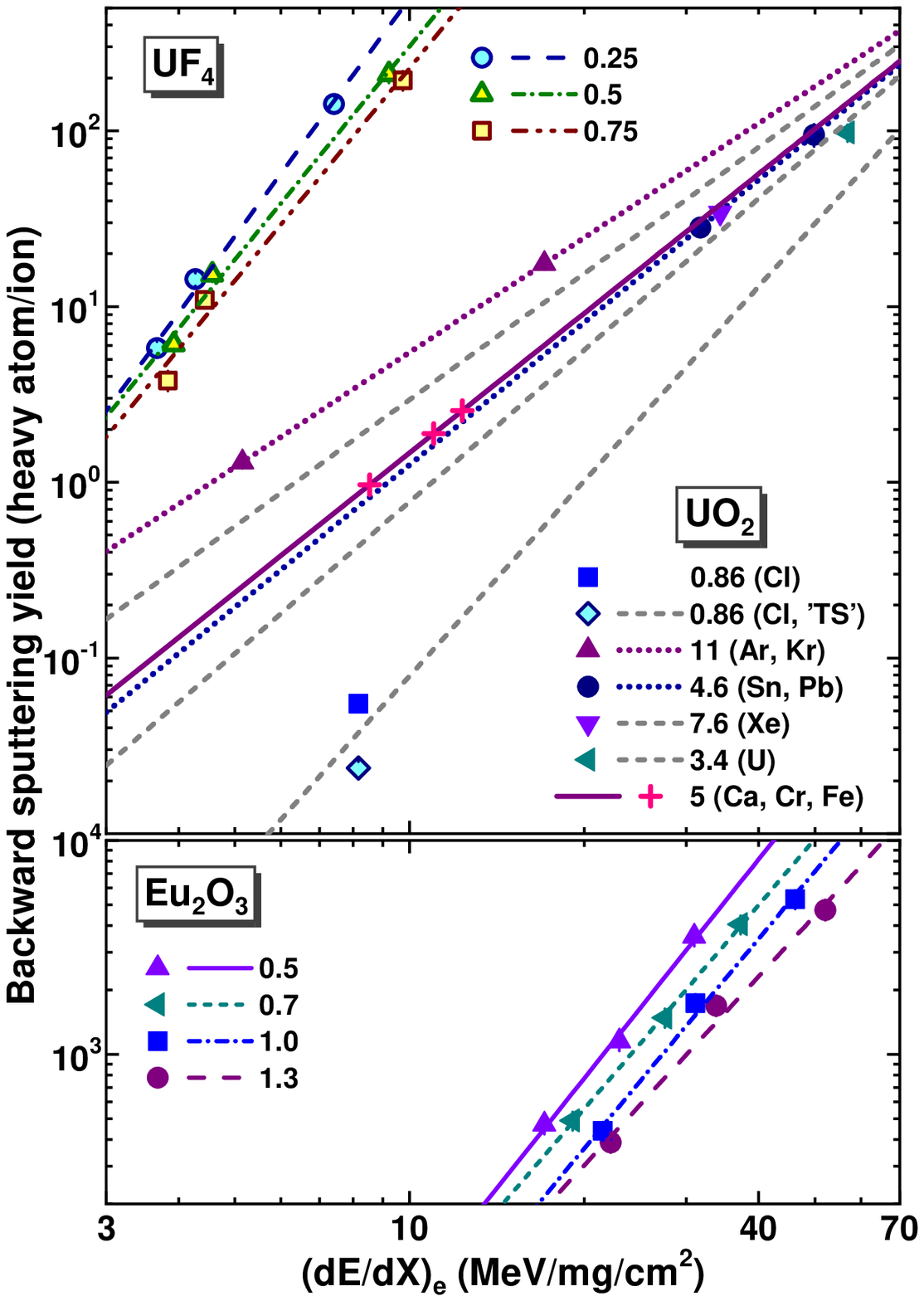}
\vspace{-1.0mm}
\caption{The same as in Figure \ref{SputdEdX}, but for the sputtering yields corresponding to the same ion velocity (symbols corresponding to the MeV/nucleon values are indicated in the figure). The results of the $a[(dE/dX)_{e}]^{b}$ function fitting the data are shown by different lines (dotted lines for the UO$_{2}$ target). The interpolated (extrapolated) results for the UO$_{2}$ data obtained in the experiments with Cl, Xe, and U ions at the respective MeV/nucleon values are shown by dashed lines in the upper panel. The same is shown for the 5 MeV/nucleon projectiles and the specified values for the $^{48}$Ca, $^{54}$Cr, and $^{58}$Fe projectiles (the solid line and crosses).}
\label{SpuVdEdX}
\end{center}
\end{figure}

Better scaling was achieved by grouping the data according to the same velocity \cite{Hedin87}, or to the same energy per nucleon, at least for the UF$_{4}$ and Eu$_{2}$O$_{3}$ targets. It is clearly seen in Figure~\ref{SpuVdEdX}, where the data points corresponding to the indicated mean energies in MeV/nucleon are shown together with the results of the $a[(dE/dX)_{e}]^{b}$ function fitting each data group. Thus, $b$ values resulting from these fits are reduced from 4.5 to 4.0 for UF$_{4}$ and from 3.4 to 2.9 for Eu$_{2}$O$_{3}$ when the HI energy increases from 0.25 to 0.75 and from 0.5 to 1.3 MeV/nucleon, respectively. The data for U atoms sputtered from the UO$_{2}$ target by $^{40}$Ar and $^{78}$Kr at the mean energy of 11.05 and by $^{116}$Sn and $^{208}$Pb at the mean energy of 4.55 MeV/nucleon \cite{Schlut01} allow us to find out the $a$ and $b$ parameter values. These parameters are changed from 0.037 to 0.0025 and from 2.17 to 2.69 for $a$ and $b$, respectively, with the corresponding energy decrease.

Fitted parameter values can be used for an interpolation (extrapolation for UF$_{4}$) to the desired HI energy per nucleon. The results of such approximations for the UO$_{2}$ target irradiated by other ions are also shown in Figure~\ref{SpuVdEdX}. The deviation of the approximated sputtering yield from the measured one for the 0.86 MeV/nucleon Cl ions (far extrapolation) is as much as 32\%, whereas similar values for Xe and U (7.6 and 3.4 MeV/nucleon, respectively) are estimated at 64\% and 16\%. One can take into account that the beam charge states for $^{35}$Cl ions \cite{Meins83} correspond to  0.61--0.67 of the equilibrated charges \cite{Ziegler80}, whereas the beam charge states for $^{40}$Ar to $^{238}$U ions \cite{Schlut01} exceed the equilibrated charges by a factor of 1.02--1.39. In general, one can assume that the accuracy of such an approximation is within a factor of two, i.e., corresponds to the accuracy typical for CCM estimates \cite{Eckstein07}. Unfortunately, because of a lack of high-energy data, a far extrapolation for the UF$_{4}$ target to the HI velocity of interest (5 MeV/nucleon) seemed doubtful and was not performed.

Within this approach and using the appropriate $(dE/dX)_{e}$ values \cite{SRIM}, the yields of U atoms sputtered from UO$_{2}$ by the 1.5--12 MeV/nucleon $^{48}$Ca beam were estimated, as an example. These data corresponding to TSM are shown in Figure~\ref{UErSput}, together with the estimates obtained with TRIM simulations. As we can see, the TSM yields occupy an intermediate position between the Cl and Ar data \cite{Meins83,Schlut01}. Note that sputtering yields obtained for the $^{48}$Ca, $^{54}$Cr, and $^{58}$Fe projectiles at the energy of 5 MeV/nucleon are lower by factors 3.5, 2.8, and 2.5, respectively (see Figure~\ref{SpuVdEdX}) than the values estimated with the same function fitting all the data for UO$_{2}$ \cite{Schlut01} (see Figure~\ref{SputdEdX}). The estimates corresponding to the definite ion velocity seem reasonable for the yields of U atoms backward sputtered from the UO$_{2}$ target, which are obtained with the $a[(dE/dX)_{e}]^{b}$ function fitting the data.

Following extrapolation (interpolation) of the fitted $a$ and $b$ parameter values to the desired ion velocity allows us to estimate the sputtering yields of U atoms under the 5 MeV/nucleon HI beams (see Figure \ref{SpuVdEdX}).  Thus, for the $^{48}$Ca beam, the backward sputtering yield is estimated at $\simeq$1 atom/ion. This estimate is about two orders of magnitude higher than the yield obtained with TRIM simulations, i.e., using the CCM approach. For transmission sputtering, which is of interest in high beam-dose experiments, this value should be multiplied by a factor of 2 if we consider extrapolation of the low-energy data for Cl ions interacting with UF$_{4}$ \cite{Meins83} (see Figure \ref{EnerDepU}). Simultaneously, TRIM simulations show about the same values for transmission and backward sputtering.  Thus, the transmission sputtering of U atoms from the UO$_{2}$ target under  the 5 MeV/nucleon $^{48}$Ca beam could be estimated as 1--2 atoms/ion. This value means that the 0.4 mg/cm$^{2}$ UO$_{2}$ stationary target should disappear for 15--30 hours at the intensity of $\sim$10$^{13}$ for the $^{48}$Ca beam. The rotating target of the same density and respective area, which was used in the experiments \cite{OgaUtRPP15}, should disappear after about 20--40 days of irradiation! Of course, it was not observed in the experiments, and we attempted to derive a reliable estimate of the sputtering yield from data obtained during the SHN campaign \cite{OgaUtRPP15}.

\section{Observations in experiments}
\label{expobs}

Attempting to find validation in the experiments for the estimates considered in the previous section, it must be noted that conditions of the irradiation of actinide targets in the SHN experiments differ significantly from those inherent in experiments on sputtering yield measurements. Thus, the incident ion beam flux has to be kept sufficiently low to avoid excessive heating, melting, or even evaporation of the sample in sputtering experiments. Practical experience suggests that the beam current density should not exceed the value of $\sim$10$^{10}$ ions/(s$\cdot$cm$^{2}$) \cite{Miesk03,AssTouTra07}. We recall that in SHN experiments with the $^{48}$Ca beam, the beam intensity was an average of 7$\times$10$^{12}$ ions/s, which corresponded to $\sim$2$\times$10$^{11}$ ions/(s$\cdot$cm$^{2}$) for the used rotating target \cite{OgaUtRPP15}. Such beam density leads to the significant target heating up to the temperature of several hundred Celsius degrees, despite the target rotation \cite{SagaPRAB21}. Note that the estimates of sputtering yields obtained above will be treated as being related to the atoms/(ion$\cdot$cm$^{2}$) values for their further comparison with experimental data.

\subsection{Experimental conditions}
\label{expcond}

As was mentioned above, a rotating target allows us to reduce the yield of sputtered atoms by increasing the irradiated area. This reduction corresponds to a factor of $\eta \simeq 8 R_{c}/d_{b}$, where $R_{c}$ is the central radius of the rotating target wheel and $d_{b}$ is the characteristic size (diameter) of a beam spot on the target surface. For example, if $d_{b}$=10 mm, the $\eta$ value is estimated at 48 and 200 for the ``small'' and ``large''  rotating targets with $R_{c}$=60 and 250 mm, respectively. These conditional estimates apply to the uniform beam-density distribution over a target surface. In the SHN experiments \cite{OgaUtRPP15}, a quasi-uniform distribution was achieved by using the ``small'' rotating target and beam wobbling in a direction perpendicular to the direction of rotation \cite{SagaPRAB21}.

Vacuum conditions are also crucial for the correct determination of sputtering yields, since contamination or oxidation of the surface can distort the measured or expected yield. A dynamically clean surface can only be maintained if the sputtering rate is much larger than the adsorption rate of contaminants. A rule of thumb gives an adsorption rate of 10$^{15}$ particles/(s$\cdot$cm$^{2}$) at a pressure of 10$^{-6}$ mbar \cite{AssTouTra07,Miesk03}. In contrast to these conditions, the rotating actinide target itself and its Ti backing are in the atmosphere of rarefied hydrogen or helium gas at a pressure of $\sim$1 Torr (a typical value for gas-filled recoil separators (GFRS) exploited for the synthesis of SHN) \cite{OgaUtRPP15,DullRev}.

In SHN high-intensity beam experiments, fission-like fragments (FLFs) produced in fission (quasi-fission) nuclear reactions induced by HI projectiles on target nuclei may contribute to the sputtering yield observed in experiments. This contribution may be non-negligible, bearing in mind that the ESPs of FLFs exceed those for HI projectiles. Indeed, at the $^{48}$Ca energy of 5 MeV above the fusion barrier for the reaction with $^{238}$U, FLFs with the most probable atomic mass numbers of 76 and 210 and a total kinetic energy of 234 MeV are produced with a cross section of 200 mb \cite{Kozu16}. Although ESPs for these FLFs are 1.5--2 times higher than the one for the $^{48}$Ca projectile, heavy FLFs with low velocities would make a negligible additional contribution to the sputtering yield produced by the beam. Light FLFs would give a sputtering yield comparable to the one produced by the beam. However, the intensity of these ion sources is estimated to be $\sim$10$^{6}$ times lower than that of the HI projectile. Thus, the contribution of FLFs to the beam sputtering yield could be neglected.

At the estimates of the ``background'' sputtering yield in experiments, one should consider one more source of target atoms sputtering. Thus, highly radioactive actinide targets are subjected to self-sputtering of target atoms, induced by energetic $\alpha$-particles and their respective low-energy $\alpha$-recoils. The contributions of these sources to the sputtering of the target atoms could be significant, for example, in experiments with the highly radioactive $^{249}$Cf target.

In the high-intensity long-term experiments on the synthesis of SHN \cite{OgaUtRPP15}, the integrity of actinide targets was periodically checked in the course of each run. These inspections were carried out with the detection of $\alpha$-activity of the radioactive target at appropriate settings of the dipole magnet and quadrupole lenses of the separator.

\subsection{Experiments  with the $^{239}$Pu target}
\label{expPu}

\begin{figure}[!h] 
\vspace{-0.5mm}
\begin{center}
\includegraphics[width=0.685\textwidth]{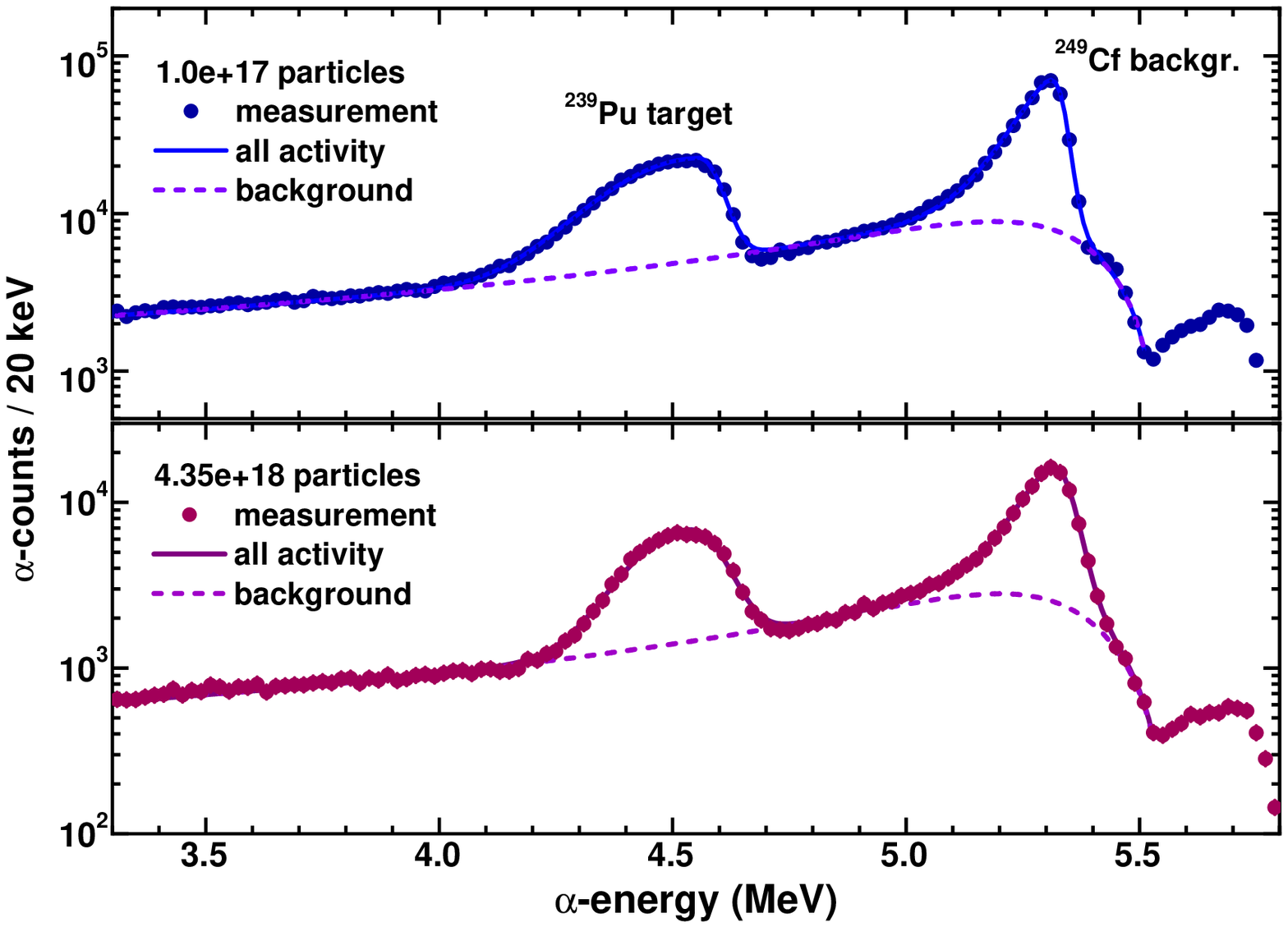}
\vspace{-0.5mm}
\caption{\label{PuCfspect}The $\alpha$-spectra from the $^{239}$Pu target as obtained at the beginning of the $^{48}$Ca+$^{239}$Pu experiment \cite{Utyon15Pu} and after collection of the significant beam dose (values are indicated in the figure). The results of fitting the spectra at $E_{\alpha}$$\leqslant$5.5 MeV and assigning background using Eq.~(\ref{sumfit}) are shown by solid and dashed lines, respectively. See details in the text.}
\end{center}
\end{figure}

In the $^{239}$Pu experiment \cite{Utyon15Pu}, the ``small'' rotating  target of the average thickness of 0.50$\pm$0.05 mg/cm$^{2}$ for $^{239}$Pu was irradiated by the 245 MeV $^{48}$Ca beam (the energy in the middle of the target layer). The target was manufactured by layered electroplating of PuO$_{2}$ onto 0.71--0.72 mg/cm$^{2}$ Ti foils from an alcohol-butyl solution and subsequent annealing. The total beam dose of the projectile passed through the target was 1.38$\times$10$^{19}$ particles. In contrast to the $^{249}$Cf target (see below), sputtering induced by $\alpha$-particles and their recoils was not considered, since the $\alpha$-activity of the $^{239}$Pu target ($T_{1/2}^{\rm Pu}$=2.41$\times$10$^{4}$ y) \cite{NuDat} was about 32 times less than the one for Cf. The Pu target irradiation was carried out using a target chamber earlier exploited in the previous Cf experiment  \cite{Ogan06Cf}. As a result, the $\alpha$-activity of Cf that sputtered around the target earlier was present in the measured $\alpha$-spectra. Thus, we were dealing with the complex $\alpha$-spectra, as shown in Figure~\ref{PuCfspect}, as an example. The prominent $\alpha$-peaks corresponding to the most intensive  $\alpha$-lines of $^{239}$Pu and $^{249}$Cf were shifted to lower energies as compared to the most intense tabulated values: 5.157 (70.8\%), 5.144 (17.1\%), and 5.106 (11.9\%) MeV for Pu, and 5.811 (82.4\%), 5.848--6.192 (12.0\%), and 5.757 (4.7\%) MeV for Cf  \cite{NuDat}. These shifts correspond to the energy losses in the stopping media of the separator, and, in addition, Pu $\alpha$-particles lost their energies in the target itself.

Target spectra obtained during irradiation were fitted using the weighted LSM for the sum of activities $A(E)$ composed of appropriate model functions:
\begin{equation}\label{sumfit}
 A(E) = A_{\rm Pu}(E) +  A_{\rm Cf}(E) + A_{b}(E),
\end{equation}
where $A_{\rm Pu}(E)$, $A_{\rm Cf}(E)$, and $A_{b}(E)$ correspond to the Pu and Cf peaks, and background present in the spectra. The Pu peak counts were described with the bi-Gaussian model function $A_{\rm Pu}(E)$:
\[ A_{\rm Pu}(E) = \left\{
\begin{array}{ll}
H_{\rm Pu} \exp\{-0.5 [(E - E_{\rm Pu}^m)/\sigma_{1}]^2\}, & \mbox{if } E < E_{\rm Pu}^m \\
H_{\rm Pu} \exp\{-0.5 [(E - E_{\rm Pu}^m)/\sigma_{2}]^2\}, & \mbox{if } E \geqslant E_{\rm Pu}^m;
\end{array} \right. \]
where $H_{\rm Pu}$ (amplitude), $E_{\rm Pu}^m$ (position of maximum), $\sigma_{1}$ and $\sigma_{2}$ (widths) are fitting parameters. The Cf peak counts were better described by the asymmetric peak function $A_{\rm Cf}(E)$:
\begin{equation}\label{asym}
  A_{\rm Cf}(E) = \frac{H_{\rm Cf} - H_{\rm Cf}/\{1 + \exp[-(E - E_{\rm Cf}^m + w_{3}/2)/w_{2}]\}}{1 + \exp[-(E - E_{\rm Cf}^m - w_{3}/2)/w_{1}]},
\end{equation}
where $H_{\rm Cf}$ (amplitude), $E_{\rm Cf}^m$ (position of maximum), $w_{1}$, $w_{2}$, and $w_{3}$ (widths) are fitting parameters. Background counts were described with the rational function $A_{b}(E)$ in the form:
\begin{equation}\label{backgr}
 A_{b}(E) = (a + b E)/(1+ c E + d E^2),
\end{equation}
where $a$, $b$, $c$ and $d$ are fitting parameters.

An example of the result of such fitting is shown in Figure~\ref{PuCfspect}, where the functions of $A(E)$ and $A_{b}(E)$, corresponding to the best fitting of spectra using $w_{3} = 0$ (the least $\chi^{2}_{r}$ value), are plotted. By integrating the fitted counts for the $A_{\rm Pu}(E)$ function, $\alpha$-activity relating to Pu was thus estimated for each spectrum.

Note that in the present approach, simplified analytical expressions, as compared to those proposed in the literature [see, for example, \cite{Garcia81,Bortels87}], were used, given that a peak-shape function in the $\alpha$-spectra measured with Si detectors differs from the Gaussian one. Such simplification was enough for the assessment of changes in relative $\alpha$-activities by reproducing the shapes of peaks in the observed $\alpha$-spectra.

\begin{figure}[!h] 
\vspace{-0.5mm}
\begin{center}
\includegraphics[width=0.685\textwidth]{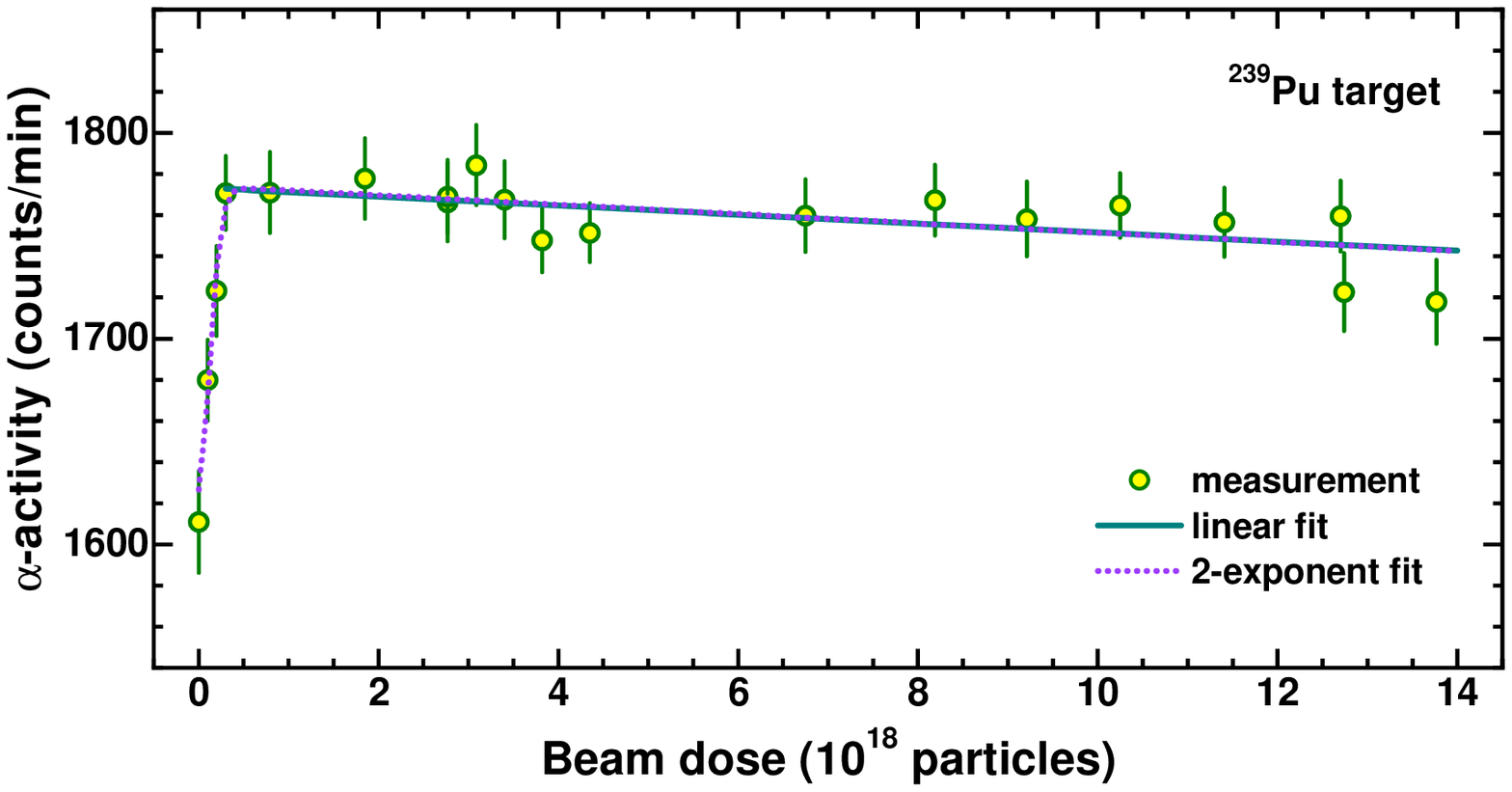}
\vspace{-17.5mm}
\caption{\label{doseactPu}The $^{239}$Pu target $\alpha$-activity as a function of the $^{48}$Ca beam dose is shown (open circles with error bars) together with the results of fitting with a linear function (starting from the fourth data point, corresponding to  the beam dose of 3$\times$10$^{17}$ particles) and with a two-exponential function applied to all data points (solid and dotted lines, respectively).}
\end{center}
\end{figure}

Figure \ref{doseactPu} shows the derived $^{239}$Pu target activity as a function of the $^{48}$Ca beam dose. The data point errors correspond to the respective errors in the parameter values of the $A_{\rm Pu}(E)$ function, as obtained with Eq.~(\ref{sumfit}), followed by integration of $A_{\rm Pu}(E)$. An increase in the activity from the very beginning of irradiation up to the beam dose of 0.3$\times$10$^{18}$particles differs from the similar dependence obtained for $^{249}$Cf, where a decrease in the activity was observed from the beginning of irradiation (see Figure~\ref{Cfactiv} below). Leaving aside the data at the beam dose less than 3$\times$10$^{17}$ particles, a linear data fitting gives us a relative value of Pu sputtered atoms corresponding to 1.69$\pm$0.74\% at the beam dose of 1.4$\times$10$^{19}$ particles. This loss corresponds to the forward sputtering yield of (4.75$\pm$2.09)$\times$10$^{-5}$ atoms/(ion$\cdot$cm$^{2}$).

\begin{figure}[!h] 
\vspace{-0.5mm}
\begin{center}
\includegraphics[width=0.685\textwidth]{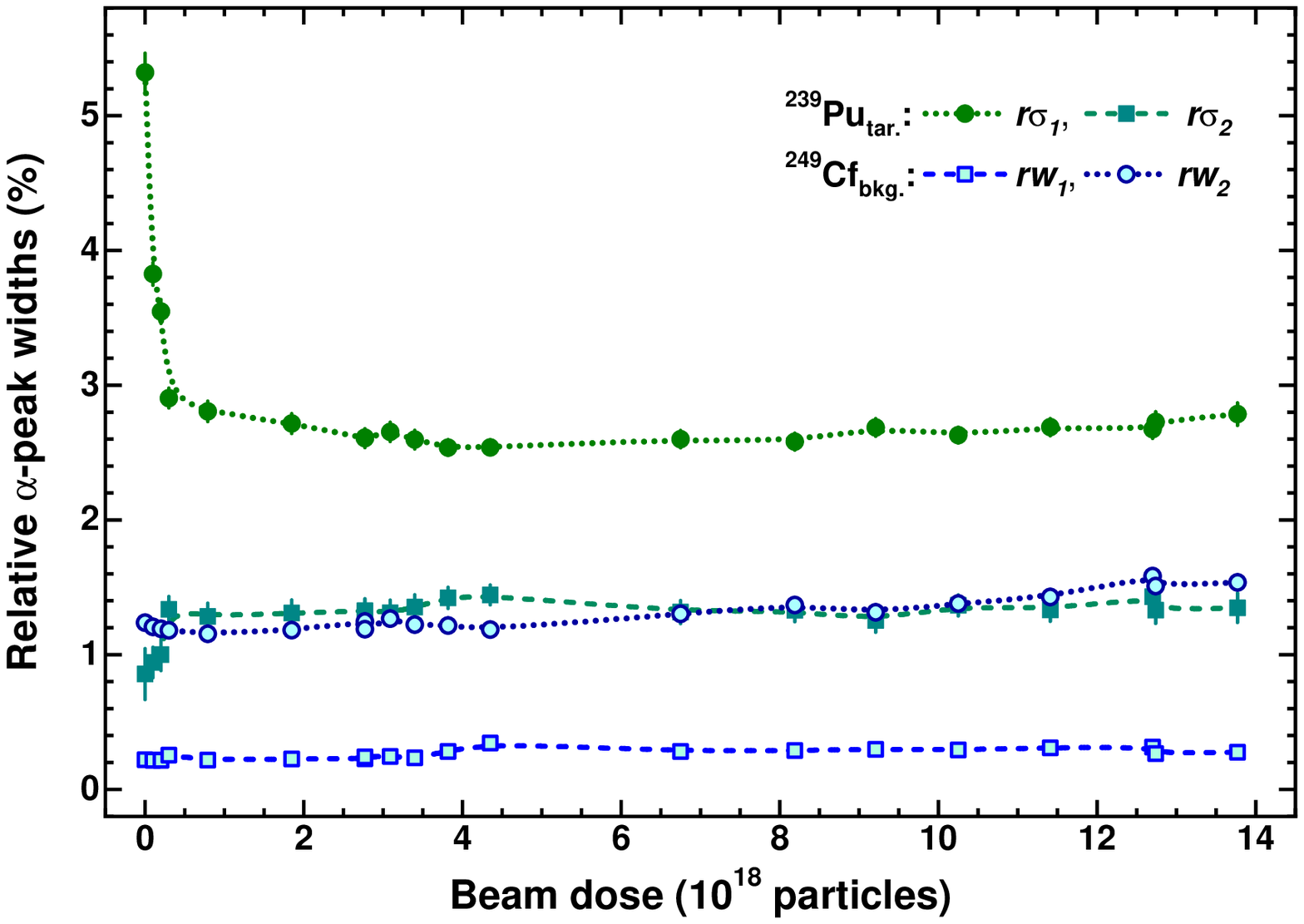}
\vspace{-0.5mm}
\caption{\label{CfPuwidths}Relative widths for the $^{239}$Pu and $^{249}$Cf peaks (closed and open symbols, respectively), as obtained by Pu target spectra fitting with Eq.~(\ref{sumfit}), are plotted against the $^{48}$Ca beam dose. Data points are connected by B-spline interpolated lines to guide the eyes.}
\end{center}
\end{figure}

An increase in the Pu activity at the very beginning of irradiation correlates with the change of relative widths $r\sigma_{1} = \sigma_{1} / E_{\rm Pu}^m$ and  $r\sigma_{2} = \sigma_{2} / E_{\rm Pu}^m$ for the respective Pu peaks. These correlations are seen in Figure~\ref{CfPuwidths}, where $r\sigma_{1}$ (determining the low energy part of peaks) drops from 5.32$\pm$0.14 to 2.90$\pm$0.07 percents, whereas $r\sigma_{2}$ (determining the high energy part of peaks) is growing from 0.86$\pm$0.19 to 1.34$\pm$0.10 percents when the beam dose increases from 0 to 3$\times$10$^{17}$ particles. At a beam dose greater than 3$\times$10$^{17}$ particles, the values of $r\sigma_{1}$ and $r\sigma_{2}$ are approximately the same. As for the similar dependencies of $rw_{1} = w_{1} / E_{\rm Cf}^m$ and  $rw_{2} = w_{2} / E_{\rm Cf}^m$ for the Cf-background peaks, which are also shown in the figure, they demonstrate a lack of dependence on the dose. It may be recalled that in the case of the Pu target, we deal with a ``thick'' $\alpha$-source, whereas the Cf source corresponds to a ``thin'' $\alpha$-emitter distributed over the inner walls of the target chamber.

\subsubsection{TRIM simulations}
\label{TRIMsim}

\begin{figure}[!h] 
\vspace{-0.5mm}
\begin{center}
\includegraphics[width=0.525\textwidth]{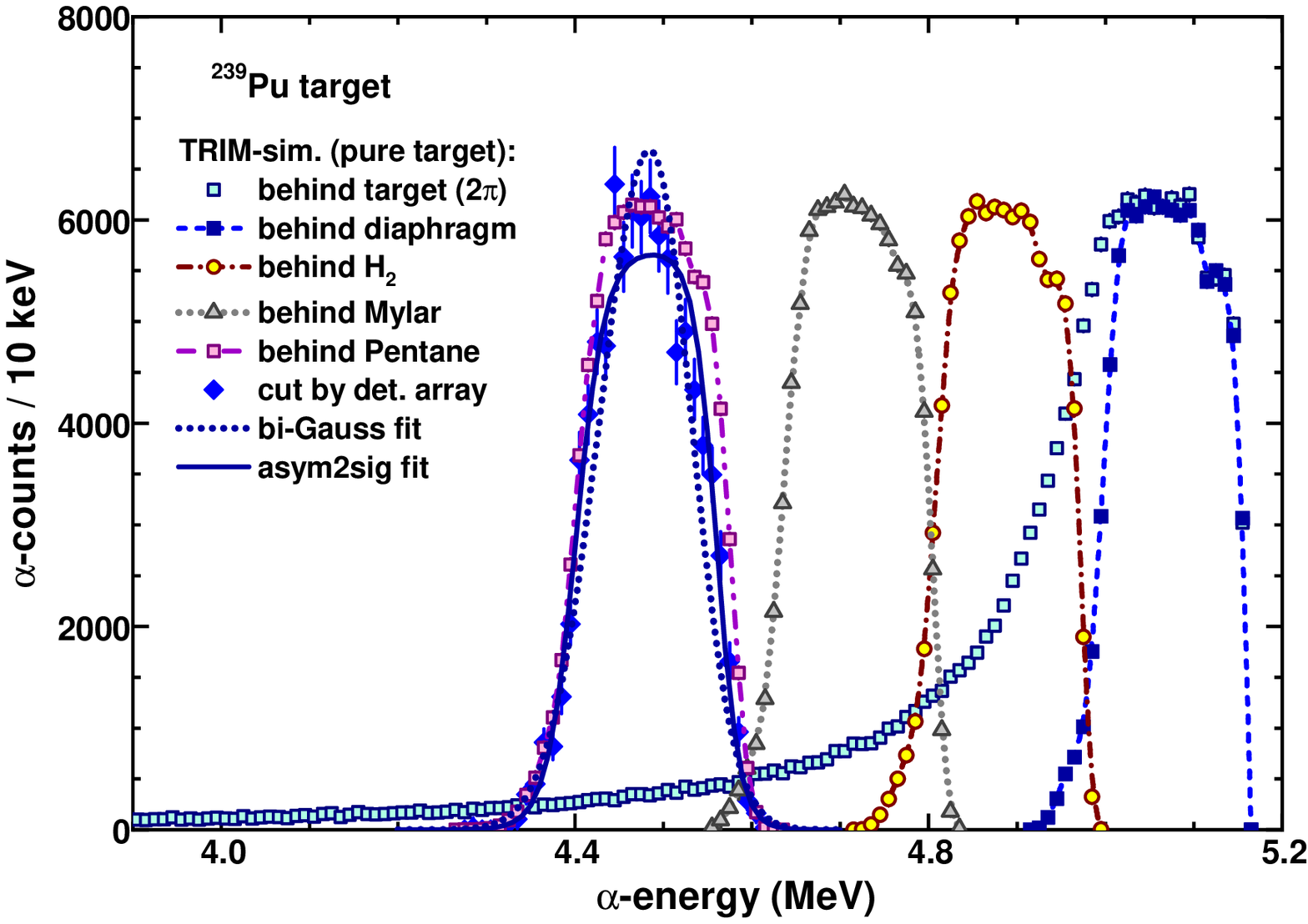}\hspace{-9.0mm}\includegraphics[width=0.525\textwidth]{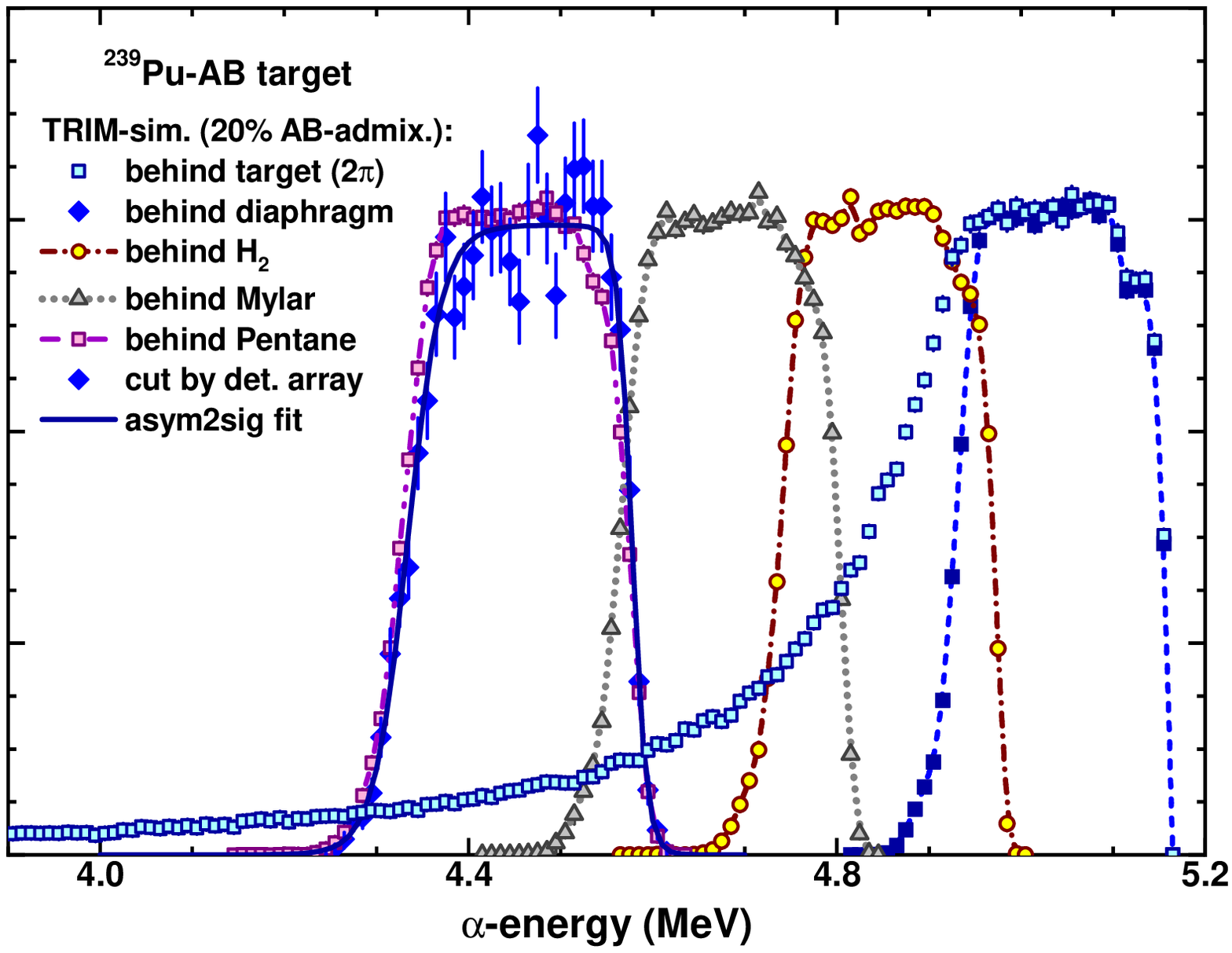}
\vspace{-0.5mm}
\caption{\label{aPuTRIM}$\alpha$-peaks obtained by TRIM simulations for the most intensive $\alpha$ lines are shown for those escaped from the pure PuO$_{2}$ target (left panel) and the same target with 20\% alcohol-butyl (AB) admixture inside (right panel), passed through the stopping media of the separator, and cut by the detector array (different symbols as indicated in the panels). The last peaks  were fitted with the Eq.~(\ref{asym}) function shown by solid lines.}
\end{center}
\end{figure}

In an attempt to reproduce the evolution of the Pu activity and peak form observed in the measurements, TRIM simulations were performed for the most intensive $\alpha$-lines emitted from the Pu target (see above). These $\alpha$-particles emitted in the forward hemisphere were slowed down in a target layer, cut by an entrance aperture of the separator, passed through the separator, and detected by an array of Si detectors at the separator's focal plane. Inside the separator, $\alpha$-particles lose ,energy in the layers of H$_{2}$ under the pressure of 1 Torr, Mylar of 1.4 $\mu$m of thickness (separates a time-of-flight (TOF) system from the GFRS volume) and pentane gas under the pressure of about 1.5 Torr (working environment of proportional counters composed of the TOF system) \cite{OgaUtRPP15}. Admixture of the alcohol-butyl (AB) remaining after target preparation was considered to be the reason for broad Pu $\alpha$-peaks at the very beginning of irradiation. The results of such simulations for the  pure target and for the one containing 20 atomic percent of AB distributed evenly over the target layer (Pu-AB target further) are shown in Figure~\ref{aPuTRIM} as an example.

\begin{figure}[!h] 
\vspace{-0.5mm}
\begin{center}
\includegraphics[width=0.525\textwidth]{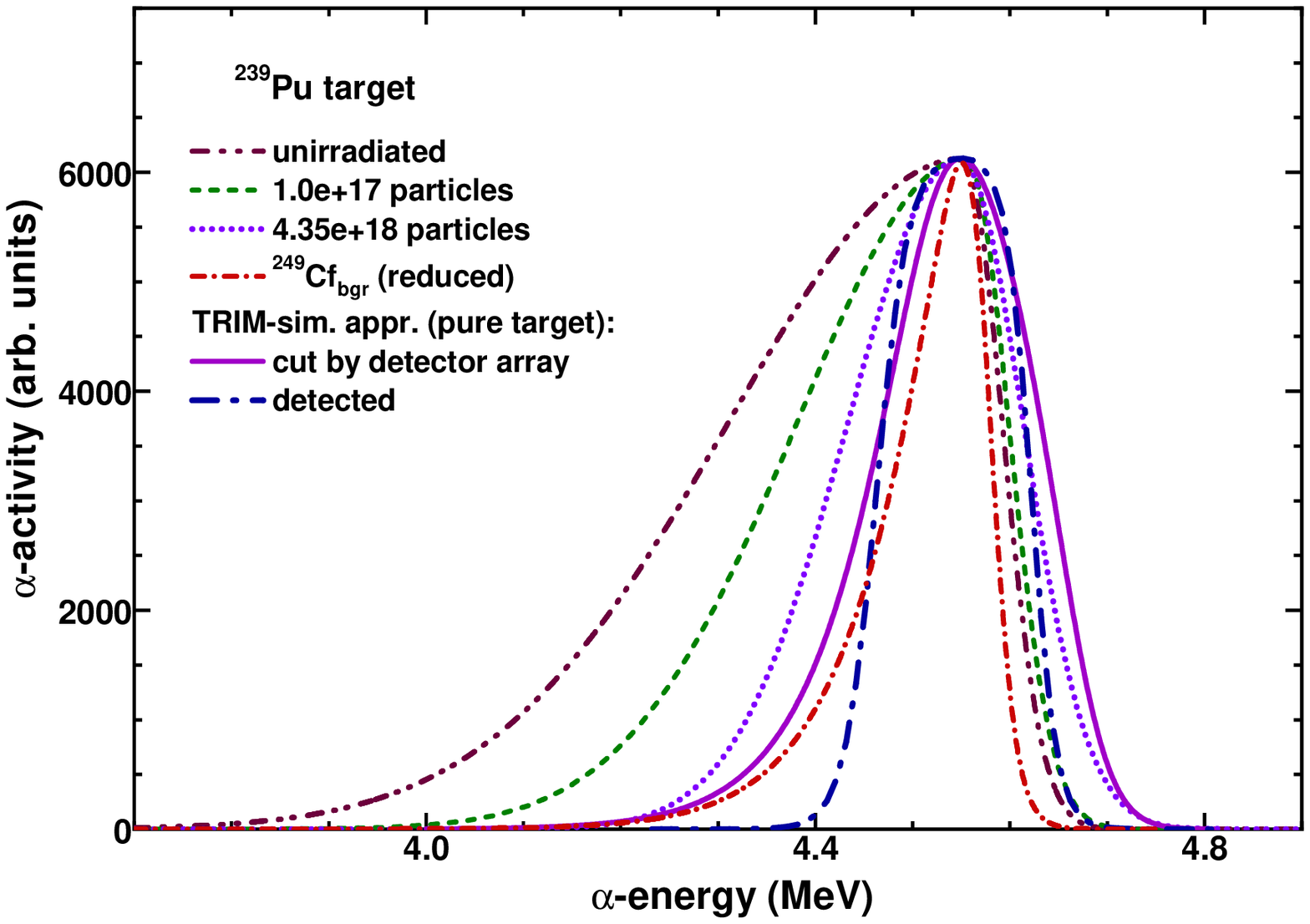}\hspace{-8.5mm}\includegraphics[width=0.525\textwidth]{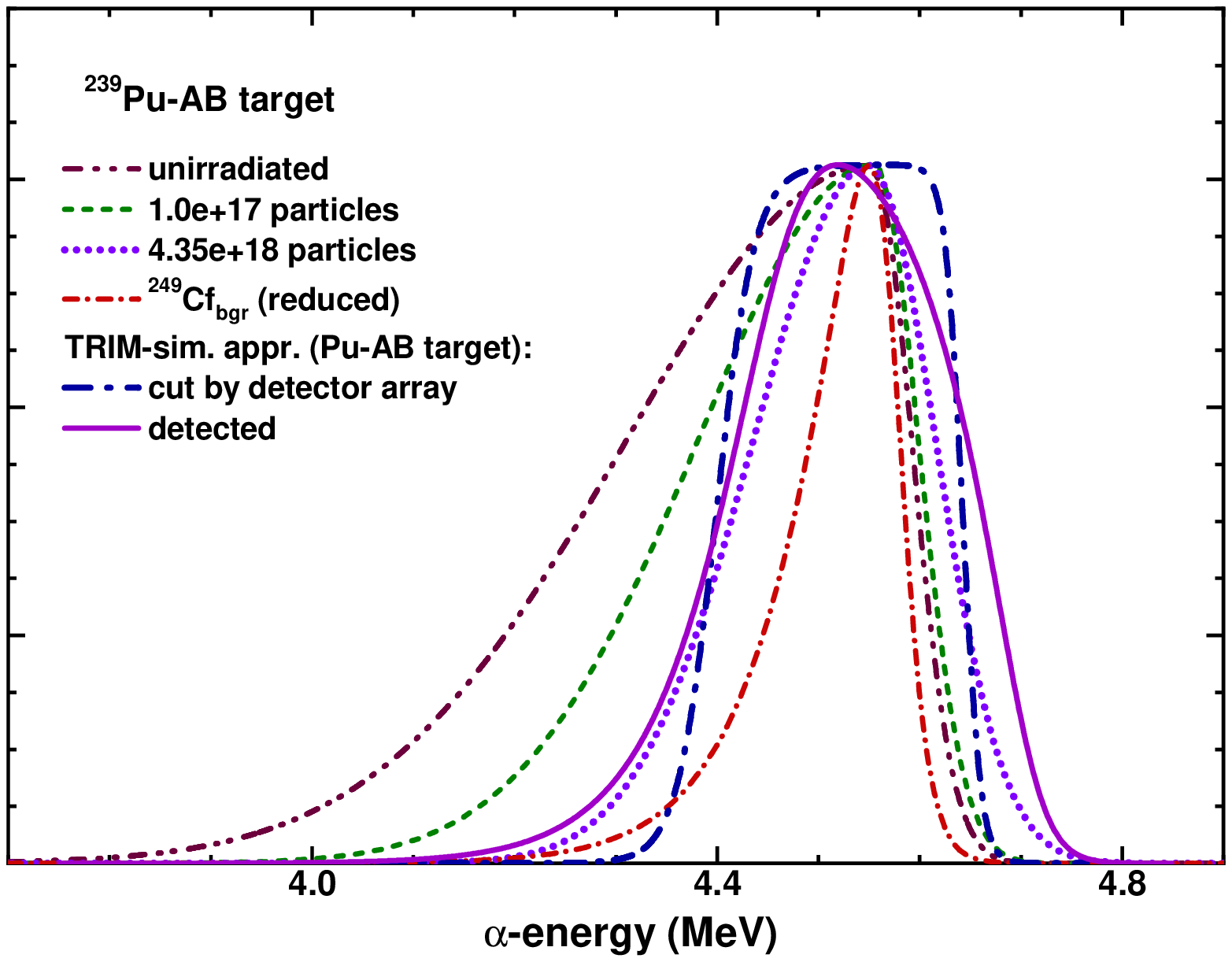}
\vspace{-0.5mm}
\caption{\label{aPuApprox}Approximated $\alpha$-peaks obtained by fitting the $\alpha$ spectra with Eq.~(\ref{sumfit}) for the unirradiated target and those measured after accumulating $^{48}$Ca beam doses of 1.0$\times$10$^{17}$ and 4.35$\times$10$^{18}$ particles are shown in both panels. These curves are compared with the TRIM-approximated curve for the pure Pu target (left panel) and the curve for the target with 20\% alcohol-butyl admixture (right panel) and with the ``detected'' curves obtained following the application of the instrumental factor using the reduced $E^{m}_{\rm Cf}$ energies of the Cf $\alpha$-peak. All curves are reduced to the same maximum value and position, except ``detected'' one for the Pu-AB target, for their better comparison to the measurements.}
\end{center}
\end{figure}

As one would expect, the TRIM-simulated peak cut by the detector array for the Pu-AB target is noticeably wider than the one for the pure PuO$_{2}$ target. For further consideration, both of these peaks were fitted with the function expressed by Eq.~(\ref{asym}), as shown in Figure~\ref{aPuTRIM} (note that fitting the bi-Gauss function showed a worse result, i.e., a noticeably larger $\chi_{r}^{2}$ value). In order to compare the detected Pu $\alpha$-peaks with those obtained in the simulations, the $\alpha$-peak form from $^{249}$Cf corresponding to the tabulated $E_{\alpha} = 5.811$ MeV converted to fitted energy $E^{m}_{\rm Cf}$ [see Figure~\ref{PuCfspect} and Eq.~(\ref{asym})] was used as an instrumental factor when detecting. In Figure~\ref{aPuApprox}, approximated curves for the detected Pu peaks obtained by fitting the $\alpha$ spectra with Eq.~(\ref{sumfit}) are shown. These curves correspond to the $\alpha$-activity of the unirradiated target and those measured after accumulating $^{48}$Ca beam doses of 1.0$\times$10$^{17}$ and 4.35$\times$10$^{18}$ particles. A comparison was made with the TRIM-approximated curve for the pure Pu target and the curve for the target with 20\% alcohol-butyl admixture. The measured curves are also compared with the TRIM-approximated curves following the application of the instrumental factor corresponding to the reduced energies of the Cf background $\alpha$-peak. These are considered the ``detected'' curves, comparable to the measured ones.

As one can see in the figure, the width of the ``detected'' curve obtained for the pure target is slightly less than a similar value derived from the measurement after reaching the beam dose of  4.35$\times$10$^{18}$ particles. Simultaneously, the width of the ``detected'' curve obtained for the Pu-AB target slightly exceeds the value derived from the same measurement. Qualitative agreement between the simulated and measured peaks might support the assumption about the presence of alcohol-butyl atoms in a freshly prepared target. This raises the question of how these atoms escape from the target with an increasing beam dose, as shown by decreasing width of peaks observed in the measurements. In trying to answer this question, TRIM simulations were performed for the estimate of the numbers of H, C, and O atoms sputtered by the 245-MeV $^{48}$Ca beam interacting with the Pu-AB target. These simulations showed that the total sputtering yield of these atoms corresponded to (2.592$\pm$0.038)$\times$10$^{-2}$ atoms/ion, which was equivalent to (8.222$\pm$0.121)$\times$10$^{-4}$ atoms/(ion$\cdot$cm$^{2}$) for the ``small'' rotating target. At the 4.35$\times$10$^{18}$ $^{48}$Ca ions passed through the target, we arrive at the relative value of 0.388$\pm$0.057\% for H, C, and O atoms, which could be sputtered from the PuO$_{2}$ target containing 20\% alcohol-butyl atoms. Cleary this value does not allow us to explain the observed thinning of the target exclusively by sputtering if the TRIM estimate is correct; otherwise, one would think the sputtering yield value would be much higher or that some other processes determine the thinning of the target.

\subsection{Experiments  with the $^{249}$Cf target}
\label{expCf}

The 0.23 mg/cm$^{2}$ $^{249}$Cf target ($T_{1/2}$=351 y) used in the experiments \cite{Ogan02Cf} emits 3.5$\times$10$^{7}$ $\alpha$/(s$\cdot$cm$^{2}$) in 4$\pi$ and the same number of the $^{245}$Cm $\alpha$-recoils with the most probable energies of 5.81 MeV and 95 keV, respectively. For half a year of such self-irradiation, the time is slightly less than the experiment duration \cite{Ogan02Cf}, and the target is subjected to a load of 5.5$\times$10$^{14}$ particles of each kind. According to \cite{Schlut01}, the 3.5 MeV $\alpha$-particles, bombarding the UO$_{2}$ target at an angle perpendicular to its surface, sputter $\lesssim$0.01 atoms/$\alpha$. The same $\alpha$-particles, which bombard the target at an angle of 30$^{\circ}$ to the surface, may sputter $\sim$0.3 atoms/$\alpha$ (the extrapolated value). The calculation for the 100 keV U ions bombarding the U target at an angle perpendicular to its surface shows a value of $\sim$15 atoms/ion \cite{Eckstein07}. TRIM simulations were performed to consider these data. Thus, for self-sputtering produced by 5.81 MeV $\alpha$-particles, a sputtering yield of 0.061 atoms/$\alpha$ was obtained. The estimate was done for U atoms sputtered from the U target, which are both the heaviest ones in SRIM/TRIM \cite{SRIM}. A similar estimate, but for the Cm $\alpha$-recoils (for U in fact), showed a sputtering yield of 0.78 atoms/recoil. It seems that both these TRIM estimates do not contradict the values obtained for outer $\alpha$-beam and backward sputtering. Thus, sputtering produced by decaying target atoms may give us a loss of $\sim$0.083\% of the initial target thickness for half a year. That is a much smaller loss than one could expect from the bombardment by the 5 MeV/nucleon $^{48}$Ca beam.

According to the estimates presented in Section~\ref{ESPdepSput}, the sputtering yield of U atoms is expected to be 1--2 atoms/ion for the 5 MeV/nucleon $^{48}$Ca beam. In the experiment on the synthesis of the 118-th element, now called Og, in the $^{48}$Ca+$^{249}$Cf reaction, the total beam dose of the projectile passed through the target was 2.48$\times$10$^{19}$ particles \cite{Ogan02Cf}. This dose corresponds to the dose density of 6.58$\times$10$^{17}$ particles/cm$^{2}$ for the ``small'' rotating target (see Section~\ref{expcond}) used in the experiment. As a result, all Cf atoms had to leave the target at the beam dose density of 5.56$\times$10$^{17}$ particles/cm$^{2}$ or two less, corresponding to 84--42\% of the total dose collected in the experiment \cite{Ogan02Cf}.

\begin{figure}[!h] 
\vspace{-0.5mm}
\begin{center}
\includegraphics[width=0.685\textwidth]{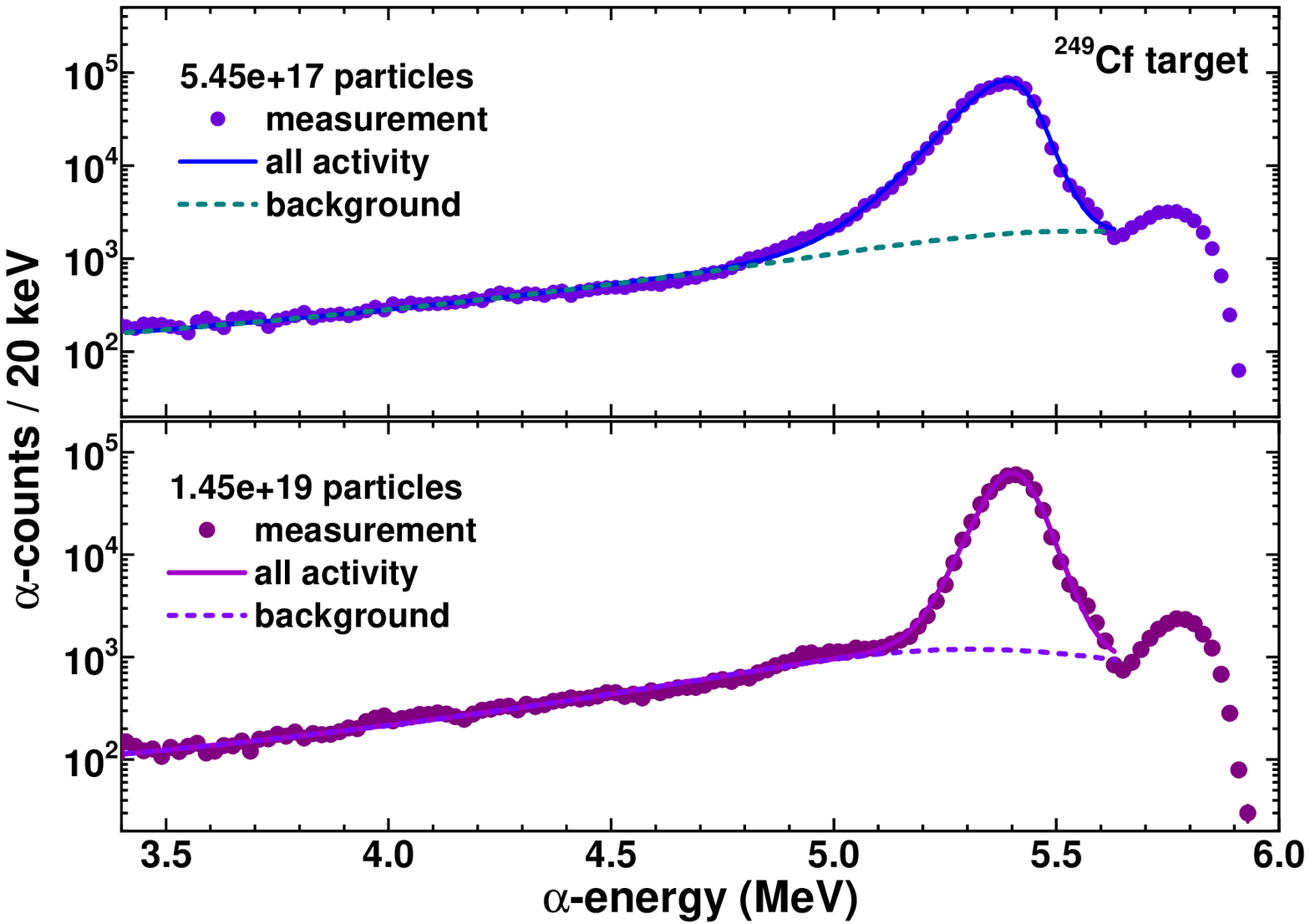}
\vspace{-0.5mm}
\caption{\label{aSpCfdose}The same as in Figure~\ref{PuCfspect}, but for the $^{48}$Ca+$^{249}$Cf experiment \cite{Ogan02Cf}. The results of fitting the most intensive $\alpha$-peak at $E_{\alpha}$$\lesssim$5.6 MeV and assigned background using Eqs.~(\ref{asym}) and (\ref{backgr}) are shown by solid and dashed lines, respectively. See details in the text.}
\end{center}
\end{figure}

The rotating $^{249}$Cf target was irradiated by the 245 MeV $^{48}$Ca beam (the energy corresponds to the middle of the target layer) \cite{Ogan02Cf}. As in the Pu-target experiment, the $\alpha$-activity of the target was periodically monitored during the run. Irradiation was carried out using a target chamber that was not exploited in previous experiments with $\alpha$-radioactive targets, i.e., background $\alpha$-activities did not prevent the measurements. Examples of $\alpha$-spectra measured after collection of the beam doses of 5.45$\times$10$^{17}$ and 1.45$\times$10$^{19}$ particles are shown in Figure~\ref{aSpCfdose}. The $\alpha$-peaks of the most intensive line, 5.811 MeV (82.4\%), and a group of 6.139--6.192 MeV (3.8\%) \cite{NuDat} were well resolved by the separator's focal plane detector. The respective energies of $\alpha$-peak maxima differ from the tabulated ones because of the energy losses of $\alpha$-particles traveling through hydrogen, Mylar foil, and pentane inside the separator, as well as in the Pu-target experiments. Variations in the Cf-target $\alpha$-activity were determined by the intensity of  the most intensive line, which, in turn, was derived by fitting measured $\alpha$-spectra using the asymmetric peak function expressed by Eq.~(\ref{asym}), taking into account the respective background described by Eq.~(\ref{backgr}).

\begin{figure}[!h] 
\vspace{-0.5mm}
\begin{center}
\includegraphics[width=0.685\textwidth]{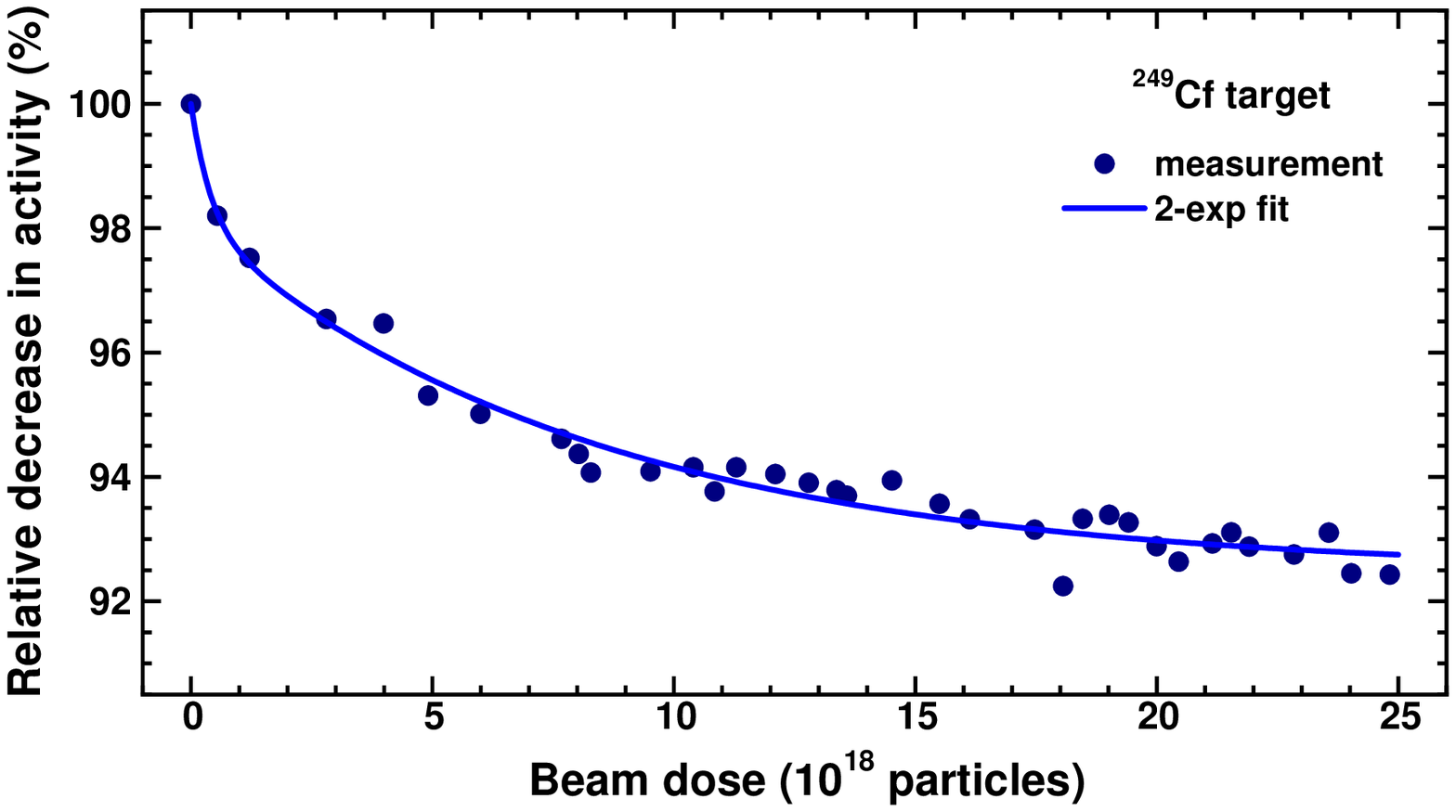}
\vspace{-14.5mm}
\caption{\label{Cfactiv}The relative $\alpha$-activity of the $^{249}$Cf target as a function of the collected beam dose, as obtained in the $^{48}$Ca+$^{249}$Cf experiment \cite{Ogan02Cf} (full circles). The two-component exponential function expressed by Eq.~(\ref{Cf2exponents}), which fits the data, is shown by a solid line.}
\end{center}
\end{figure}

In Figure ~\ref{Cfactiv}, the relative value of $\alpha$-activity is shown as a function of the $^{48}$Ca beam dose for particles passed through the target. Note that a rapid decrease in the activity of the Cf target is not similar to the linearly increasing total sputtering yield plotted as a function of the beam fluence, which is usually observed in a sputtering experiment [see, for example, \cite{Matsu13}]. The difference is that at the beginning of the irradiation, corresponding to relatively small beam doses, sputtering of Cf atoms arises more intensively than it occurs at significantly higher doses. The relative decrease in the $\alpha$-activity ($rA$ in percent) with the beam dose ($bd$ in 10$^{18}$ particle units) was approximated by the two-component exponential function in the form:
\begin{equation}\label{Cf2exponents}
  rA = 100 - a_{1} [1 - \exp(-bd/k_{1})] - a_{2} [1 - \exp(-bd/k_{2})],
\end{equation}
where $a1$, $k_{1}$, $a2$, and $k_{2}$ are fitting parameters relating to the ``fast'' and ``slow'' sputtering components.

The approximation with Eq.~(\ref{Cf2exponents}) was used for the estimates of the relative loss of target atoms at the beam dose of 2.5$\times$10$^{19}$. This value corresponds to the value of (7.25$\pm$0.55)\% Cf atoms, which value leads to the transmission sputtering yield, corresponding to  $S_{tr}$=(1.61$\pm$0.12)$\times$10$^{-3}$ atom/ion. The last value is equivalent to (5.12$\pm$0.39)$\times$10$^{-5}$ atom/(ion$\cdot$cm$^{2}$) for the rotating target. The yield of Cf atoms obtained in the experiment with the $^{249}$Cf target under the 245 MeV $^{48}$Ca beam \cite{Ogan02Cf} is much lower than the approximated one using the UO$_{2}$ target data \cite{Schlut01}, as described in Section~\ref{ESPdepSput}, and the one obtained with TRIM simulations. Simultaneously, the derived Cf sputtering yield is in a good agreement with the value obtained in the Pu-target experiment [(4.75$\pm$2.09)$\times$10$^{-5}$ atoms/(ion$\cdot$cm$^{2}$)].

\begin{figure}[!h] 
\vspace{-0.5mm}
\begin{center}
\includegraphics[width=0.685\textwidth]{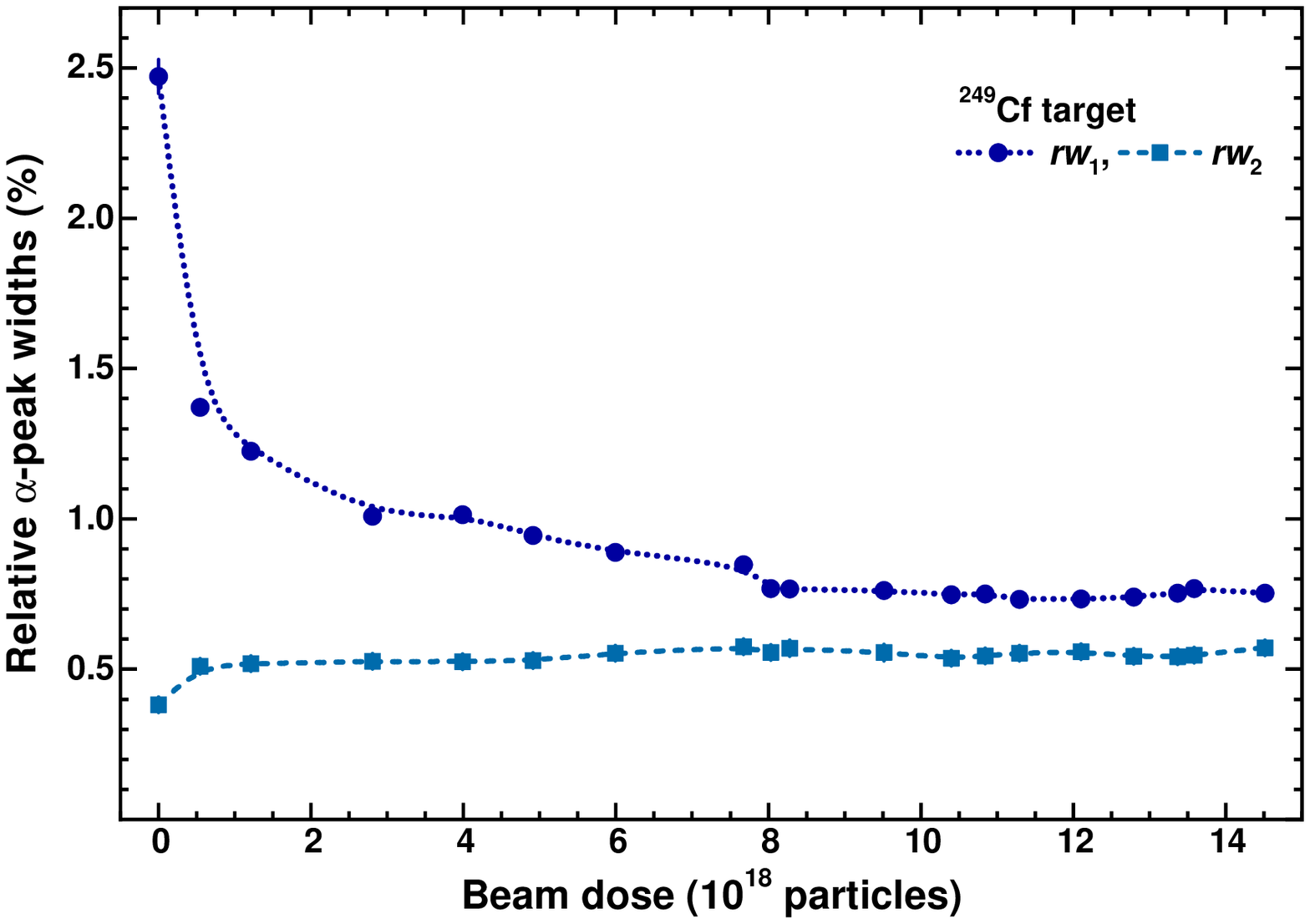}
\vspace{-0.5mm}
\caption{\label{Cfwidths}The same as in Figure~\ref{CfPuwidths}, but for the relative widths of the Cf $\alpha$-peak, as obtained by fitting Cf-target spectra with the use of Eqs.~(\ref{asym}) and (\ref{backgr}).}
\end{center}
\end{figure}

As in the case of the Pu-target experiment, a similar evolution of widths of the Cf-target $\alpha$-peak with increasing a beam dose was observed. In Figure~\ref{Cfwidths}, this evolution is shown for the relative values  $rw_{1} = w_{1} / E_{\rm Cf}^m$ and  $rw_{2} = w_{2} / E_{\rm Cf}^m$. Note that these values are significantly smaller than similar ones obtained for Pu-target $\alpha$-peaks. Such a difference could be expected, bearing in mind that a thickness of the Cf target was more than two times less than the Pu one. Despite the similarity in the behavior of widths with the beam dose, the initial changes in the $\alpha$-activity are dramatically different. TRIM simulations show that the initial broadening observed for both Pu and Cf $\alpha$-peaks may be due to the residual impurities of alcohol-butanol atoms (both targets were prepared from the respective solution). However, in the case of the rather thick Pu target, AB-admixture reduced its initial  $\alpha$-activity, whereas this admixture inside the rather thin Cf target did not reduce its initial  $\alpha$-activity. Moreover, this activity was even higher than the one obtained during subsequent measurements. The two-component $\alpha$-activity observed in the Cf-target measurements implies the presence of the high sputtering-yield component at the beginning of irradiation, which disappears in consequence. On the other hand, for the thicker Pu target with AB-admixture, a seemingly higher multiple scattering should reduce the $\alpha$-particle yield measured within a very narrow solid angle. Unfortunately, TRIM simulations do not support this assumption, at least for the 20\% AB-admixture.

\section{Summary}
\label{summa}

The synthesis and study of the properties of superheavy nuclei (SHN) with $Z$$>$118, produced in fusion-evaporation reactions induced by heavy ion (HI) projectiles, imply the beam dose collection of up to $\sim$10$^{20}$ particles passed through the target. Typical thicknesses of actinide targets in such experiments using recoil separators correspond to $\sim$10$^{18}$ atoms/cm$^{2}$. TRIM simulations \cite{SRIM} show that the yield of U atoms sputtered by the 5 MeV/nucleon $^{48}$Ca beam from the UO$_{2}$ target corresponds to $\sim$0.01 atom/ion. It means that toward the end of the experiment carried out with the collection of such beam dose, just over 3\% of actinide target atoms may leave a rotating target, similar to the one used in the discovery experiments \cite{OgaUtRPP15}. Nevertheless, the question of sputtering yields under the $^{48}$Ca to $^{58}$Fe beams at the energy of $E$ = 5--6 MeV/nucleon remains, bearing in mind that TRIM simulations, as the variant of the collision cascade model (CCM), underestimate essentially the sputtering yields at these energies \cite{AssTouTra07}.

The analysis of experimental data on the sputtering of oxides and fluorides showed underestimates corresponding to several orders of magnitude depending on $Z$ and $E$ of heavy ions. The only exception was obtained for the yields of U atoms sputtered from UO$_{2}$ by Cl ions at energy $E<1$ MeV/nucleon \cite{Meins83}. These data are in close agreement with TRIM simulations. At the same time, all the data subjected to analysis enabled an approximation with a power function of electronic stopping power $(dE/dX)_{e}$, using the relationship $S \sim [(dE/dX)_{e}]^4 = [A Q_{eq}^{2}(E,Z_{HI}) \ln(B E) / E]^4$ proposed in the framework of the thermal spike model \cite{Meins83,Seiber80}. Unfortunately, a lack of data did not allow us to systematize the fitted $A$ and $B$ parameters. However, the data on sputtering yields obtained at about the same HI velocity allowed their approximation with the $a[(dE/dX)_{e}]^{b}$ function. Such an approximation was performed for U atoms sputtered from UO$_{2}$ by different HIs at 3.4 to 11.3 MeV/nucleon \cite{Schlut01}. The interpolation of the fitted $a$ and $b$ parameter values to the desired velocity allowed one to estimate the desired sputtering yields. Thus, for transmission sputtering of U atoms from the UO$_{2}$ target under the 5 MeV/nucleon $^{48}$Ca beam, the yield was estimated as 1--2 atoms/ion. This estimate is about two orders of magnitude as high as the one obtained with TRIM simulations.

The yield of Pu atoms sputtered from the rotating $^{239}$PuO$_{2}$ target by the 245 MeV $^{48}$Ca beam was obtained during the experiment on the synthesis of Fl isotopes (of the 114th element) \cite{Utyon15Pu}, using periodic measurements of the target $\alpha$-activity. The transmission sputtering yield was estimated as $S_{tr}$=(4.75$\pm$2.09)$\times$10$^{-5}$ atoms/(ion$\cdot$cm$^{2}$). This value is much lower than the one [1--2 atoms/(ion$\cdot$cm$^{2}$)] estimated by interpolation of the UO$_{2}$ target data \cite{Schlut01}, the value of 0.01 atom/(ion$\cdot$cm$^{2}$) obtained with TRIM simulations for the same target.

The yield of Cf atoms sputtered from the rotating $^{249}$Cf target by the same beam was obtained during the experiment on the synthesis of the 118-th element \cite{Ogan02Cf}, using the same measurements. The transmission sputtering yield was estimated at  (5.12$\pm$0.39)$\times$10$^{-5}$ atom/(ion$\cdot$cm$^{2}$) for the rotating target. As well as for Pu atoms, the yield of Cf atoms is much lower than the approximated one from the UO$_{2}$ data \cite{Schlut01} and the one obtained with TRIM simulations. Simultaneously, the thus derived Cf sputtering yield is in good agreement with the value obtained in the Pu-target experiment [(4.75$\pm$2.09)$\times$10$^{-5}$ atoms/(ion$\cdot$cm$^{2}$)].

The data obtained in sputtering experiments imply that the yields of atoms are independent of the temperature of the targets. Simultaneously, our experimental data refer to relatively high temperatures confirmed by the presence of temper colors on the surfaces of the Ti target backing. The temperatures achieved in experiments with intense HI beams and rotating targets were estimated to be  several hundreds of Celsius degrees \cite{SagaPRAB21}. Thus, data on sputtering yield for the Cf and Pu atoms can be associated with these temperatures. In similar experiments with an intensity of 7$\times$10$^{13}$ s$^{-1}$ \cite{DmitrDC280} one can expect a much higher target temperature if the same ``small'' rotating target wheel is employed. In this sense, the question remains about the applicability of the extrapolated values for the sputtering yields obtained at normal temperatures to the actinide oxide targets heated to several hundreds of degrees Celsius. Data on sputtering yields as a function of temperature are required to answer this question. It is also essential to learn whether the sputtering yields of different actinide oxides are so different as we derived for the CfO$_{1.5-2}$ and PuO$_{2}$ targets, comparing them with  UO$_{2}$ data obtained earlier in experiments.

\section*{Acknowledgements}

The author  would like to express his appreciation to Dr. V.K. Utyonkov, who kindly provided the raw data from the $\alpha$-activity measurements, and to Dr. A.V. Sabelnikov, who introduced the author to the processes of actinide target preparation.

\end{document}